\documentclass[journal]{IEEEtran}

\usepackage{amsmath}
\usepackage{cite}
\usepackage{mathtools}

\usepackage{multicol}
\usepackage{soul}
\ifCLASSINFOpdf
\usepackage{graphicx}
\usepackage{multirow}
\usepackage{txfonts}
\usepackage{dblfloatfix}
\else
\usepackage{graphicx}
\fi
\usepackage{float}
\usepackage{color}
\ifCLASSOPTIONcompsoc
\usepackage[caption=false,font=footnotesize,labelfont=sf,textfont=sf]{subfig}
\else
\usepackage[caption=false,font=footnotesize]{subfig}
\fi
\usepackage{cite}
\usepackage{dblfloatfix}
\usepackage{algorithm}
\usepackage{algpseudocode}
\usepackage{breqn}

\usepackage{float}
\usepackage{makecell}
\usepackage{adjustbox}

\usepackage{caption}

\usepackage{lineno,hyperref}
 \usepackage{float}
\usepackage{multicol}
\usepackage{lipsum}
\usepackage{array}
\usepackage[acronym]{glossaries}
\usepackage{hhline}
\usepackage{url}
\usepackage{dblfloatfix}
\usepackage{lipsum}
\usepackage{fixltx2e}
\usepackage{tabu}

\usepackage{hyperref}
\hypersetup{
    colorlinks=true,
    linkcolor=black,
    filecolor=black,      
    urlcolor=black,
}
 
\begin{document}

\definecolor{amber(sae/ece)}{rgb}{1.0, 0.49, 0.0}
\definecolor{aqua}{rgb}{0.0, 1.0, 1.0}

\title{Design and Performance Analysis of Hardware Realization of 3GPP Physical Layer for 5G Cell Search}

\author{Khalid Lodhi, Jayant Chhillar, Sumit J. Darak and Divisha Sharma
					\thanks{This work is supported by the funding received from core research grant (CRG) awarded to Dr. Sumit J. Darak from DST-SERB, GoI.}
			\thanks{All authors are with Electronics and Communications Department, 
				IIIT-Delhi, India-110020 (e-mail: \{khalid17160, jayant17154, sumit, divisha19204\}@iiitd.ac.in}
	
}
	\maketitle
\begin{abstract}
5G Cell Search (CS) is the first step for user equipment (UE) to initiate the communication with the 5G node B (gNB) every time it is powered ON. In cellular networks, CS is accomplished via synchronization signals (SS) broadcasted by gNB.  5G 3rd generation partnership project (3GPP) specifications offer a detailed discussion on the SS generation at gNB but a limited understanding of their blind search, and detection is available. Unlike 4G, 5G SS may not be transmitted at the center of carrier frequency and their frequency location is unknown to UE. In this work, we demonstrate the 5G CS by designing 3GPP compatible hardware realization of the physical layer (PHY) of the gNB transmitter and UE receiver. The proposed SS detection explores a novel down-sampling approach resulting in a significant reduction in complexity and latency. Via detailed performance analysis, we analyze the functional correctness, computational complexity, and latency of the proposed approach for different word lengths, signal-to-noise ratio (SNR), and down-sampling factors. We demonstrate the complete CS functionality on GNU Radio-based RFNoC framework and USRP-FPGA platform. The 3GPP compatibility and demonstration on hardware strengthen the commercial significance of the proposed work.
\end{abstract}


\section{Introduction}
\label{Sec:introduction}
In a cellular network, user equipment (UE) performs initial access (IA) to establish the communication with the base station every time it is powered ON \cite{CSreview,IA1,IA2,IA3,IA4,IA5}. The IA comprises the downlink and uplink synchronizations. The downlink synchronization involves the cell search (CS), and acquisition of minimum system information (MSI) at the UE \cite{IA1,SS3,SS1,SS2}. The first step, CS, allows the UE to obtain cell identity and synchronize with the BS in terms of symbol, slot, subframe, and frame timings \cite{SS1,SS2,SS3}. After the CS, MSI acquisition provides information such as access type (barred, restricted, or unrestricted), carrier frequency and bandwidth, cell selection information (minimum receiver level), scheduling information, downlink/uplink configurations, etc \cite{IA1,5GBook_IA}.
The uplink synchronization, via physical random access channel (PRACH), allows the base station to locate and instruct the UE to fine-tune the uplink timings such that uplink transmissions from multiple UEs are aligned in time irrespective of their distances from base-station \cite{5GBook_IA,PRACH1,PRACH2}.  

The CS is the first and foremost step of the 5G IA and the focus of the work presented in this paper. It is accomplished via synchronization signals (SS) broadcasted by 5G node B (gNB).  5G 3rd generation partnership project (3GPP) specifications offer a detailed discussion on the SS generation at gNB but limited understanding about their blind search, and detection at UE is available \cite{SS1,SS2,5GBook_IA,SS3,3GPP_211,3GPP_201,3GPP_213}. Historically, it is left to be designed by equipment manufacturers.

The CS in 5G is substantially evolved from 4G Long Term Evolution (LTE) \cite{LTE_SS,5GBook_NR,5GBook_LTE,5GBook1}. The major difference is the location of SS in the frequency domain. In 4G LTE, the SS is transmitted over the subcarriers located in the middle of the carrier \cite{LTE_SS}. In 5G, the location is not fixed and is unknown which means the UE needs to blindly search the SS over the entire carrier bandwidth \cite{5GBook_IA}. To simplify the blind search, 3GPP introduced Global Synchronization Channel Number (GSCN) in 5G which pre-defines the number of possible positions over which gNB can transmit SS. The SS bandwidth and overall transmission bandwidth in 5G are significantly wider and hence, SS detection requires multiple large-size correlators \cite{IA3}. This makes the blind search computationally complex and time-consuming. Other blocks of the CS physical layer (PHY) including message generators, channel encoders, data modulators, scramblers, interleavers, and their counterparts in the receivers are redesigned in 5G \cite{5GBook_PHY,OFDM5}. 

In this paper, we focus on the 5G CS PHY  based on the Release 16 of the 3GPP standards \cite{3GPP_211,3GPP_201,3GPP_213}. We develop and integrate various building blocks of the 5G CS PHY of the gNB transmitter and UE receiver for realization on hardware. Along with the conventional baseband PHY operations, we design the gNB scheduler to broadcast the SS signals as per the 3GPP specifications. Next, we demonstrate the design and implementation of signal processing blocks such as primary SS (PSS) and secondary SS (SSS) detection, demodulation reference signal (DMRS) detection, and cell identity (CI) estimation at UE. The proposed PSS detection explores a novel down-sampling approach resulting in a significant reduction in complexity and latency. Multiple instances of the detected PSS, SSS, and DMRS signals are used for frame, sub-frame, and symbol boundaries synchronization.  Via detailed performance analysis, we analyze the functional correctness, computational complexity, and latency of the proposed approach for different word lengths, signal-to-noise ratio (SNR), and down-sampling factors. 

We demonstrate the functionality of the proposed 5G CS PHY on the GNU Radio and Universal Software Radio Peripheral (USRP) based radio frequency network-on-chip (RFNoC) platform from Ettus Research \cite{RFNOC1,RFNOC2}. We demonstrate the gain in execution speed by efficiently utilizing the Field-Programmable Gate Array (FPGA) available on USRP compared to conventional GNU Radio-based software implementation. The proposed demonstration on hardware and 3GPP compatibility strengthens the commercial significance of the proposed work. Please refer here\footnote{https://tinyurl.com/A2AIIITD} for source codes and supporting tutorials. 

	The rest of the paper is organized as follows. We discuss the 3GPP specifications for 5G CS and review the relevant works in Section~\ref{Sec:review}. The downlink transmitter and receiver physical layer architectures are presented in Section~\ref{Sec:DLphy} and Section~\ref{Sec:ULphy}, respectively. The experimental results and complexity analysis is done in Section~\ref{Sec:results}. In Section~\ref{Sec:RFNOC}, we discuss the realization of CS using GNU Radio based RFNoC framework followed by experimental results. Section~\ref{Sec:Conc} concludes the paper.

\section{Specifications of 5G CS and Literature Review}
\label{Sec:review}
In this section, we discuss the 3GPP specifications for 5G CS and review the works related to hardware mapping of CS PHY. 

\subsection{5G CS PHY}
We consider the n78 frequency band (3300 MHz - 3800 MHz) in frequency range 1 i.e. FR1 (410 MHz - 7125 MHz) as it is being widely chosen for initial deployment of the 5G networks \cite{IA4,IA5}. As per the 3GPP specifications, the maximum transmission bandwidth in the n78 band is 100 MHz, and sub-carriers (SC) spacing (SCS) is fixed to 30 kHz for SS \cite{3GPP_201}. Thus, the maximum number of SC in the n78 band is 3276. In 5G, a resource block (RB) is the smallest bandwidth unit for resource allocation and 1 RB consists of 12 SCs. Thus, the maximum number of RBs available in the n78 band is 273. 

The 5G PHY is based on orthogonal frequency division multiplexing (OFDM) based waveform modulation \cite{5GBook_PHY}. Each OFDM symbol consists of the number of SCs which must be a power of two due to IFFT/FFT operations. This results in 4096 SCs per OFDM symbol in n78 \cite{3GPP_201}. In the time domain, each frame duration is 10 $ms$ and it consists of 10 sub-frames. Each sub-frame consists of 2 slots and each slot consists of 14 OFDM symbols. Then, a single frame of 10 ms comprises 280 OFDM symbols. Then, the symbol duration is 33.33 $\mu$s with a cyclic prefix size of 2.86 $\mu$s for the first symbol of each slot and  2.34 $\mu$s for the remaining 13 symbols. Different size of cyclic prefix allows a common frame structure for multiple SCS newly introduced in 5G \cite{IA4,5GBook_PHY}.

The SS signals are transmitted in a burst, known as SS burst (SSB), and the size, duration, and periodicity of the SSB depend on the operating frequency range. For n78, SSB is transmitted in every other frame and each SSB comprises of 8 SS \cite{3GPP_211,3GPP_213,SS1,SS2,SS3,5GBook_IA}. In a frame, the starting OFDM symbol of the SS is \{4,8,16,20,32,36,44,48\}. The duration of each SS is 4 OFDM symbols which means SSB spans over 52 OFDM symbols in a frame of 280 symbols.

\begin{figure}[!t]
\centering
\includegraphics[width=\columnwidth]{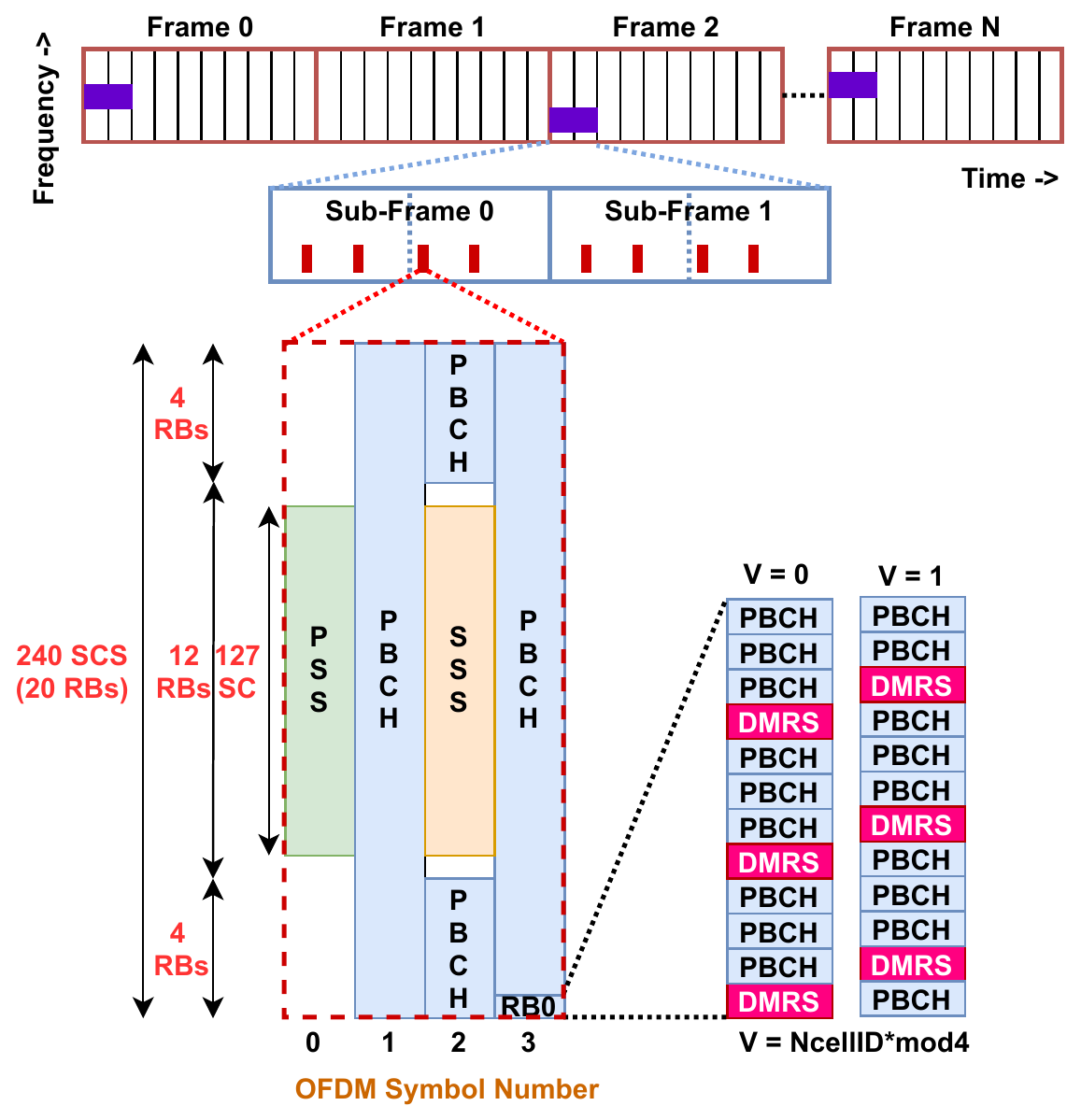}
\caption{Details of 5G SSB scheduling and SS build blocks \cite{3GPP_211,3GPP_213}.}
\label{fig:SSB}
\end{figure}

Each SS consists of 240 sub-carriers (SC) in frequency domain i.e. 20 resource blocks (RBs) and 4 OFDM symbols in time domain \cite{3GPP_211,3GPP_213,SS1,SS2,SS3,5GBook_IA}. 
As shown in Fig.~\ref{fig:SSB}, the middle 127 SC of the first and third OFDM symbols are occupied by PSS and SSS, respectively.  Since PSS and SSS are the first signals detected by UE during IA, they are carefully designed to enable blind detection with high reliability. 3GPP has adopted m-sequences to generate PSS $d_{PSS}(n)$ and SSS $d_{SSS}(n)$, $0\leq n < 127$ \cite{3GPP_211}. The rest of the 113 SCs of the first symbol are fixed to zero. The PSS and SSS allow the UE to detect the PCI. In 5G, there are 3 candidate PSS sequences and 336 candidate SSS sequences which means the PCI range is from 0 to 1007 \cite{3GPP_211,3GPP_213}. In the third symbol, there are 7 upper and 6 lower SC adjacent to the SSS fixed to zero and the remaining SCs are occupied by a physical broadcast control channel (PBCH) carrying MIS and DMRS.  The DMRS occupies 25\% of the remaining SC i.e. 144 SC. Since DMRS is QPSK demodulated, it consists of 288 bits which are generated using 3GPP defined pseudo-random sequence generator using two parameters: 1) PCI, $N_{ID}^{Cell}$, and 2) SS index, $SS_i$, of the SS in SSB, $SS_i$ $i\in\{1,2,..,8\}$ \cite{3GPP_211,3GPP_213}.
At UE, DMRS is detected after PSS and SSS detection which means DMRS allows estimation of SS index which is critical for a frame, sub-frame, slot, and symbol synchronization. DMRS also allows channel estimation for MIS reception via PBCH. 


The location of all SS in SSB among 4096 SCs is not fixed in 5G and it is decided by the GSCN value. Furthermore, gNB can dynamically change the GSCN of the SSB. As per 3GPP specifications, there is a one-to-one mapping between the GSCN value and corresponding SSB center frequency \cite{3GPP_211,3GPP_213}. In n78, the GSCN range is from 7711 - 8051 and the frequency resolution is 1.44 MHz. For example, the carrier frequency of 3305.28 Mhz is assigned a GSCN of 7711. Then, the GSCN of 7712 corresponds to the carrier frequency of 3305.28 + 1.44 = 3306.72 MHz. The last GSCN of the n78 band is 8051 and it corresponds to 3794.88 MHZ \cite{3GPP_211,3GPP_213}. Such a GSCN raster of 1.44 MHz resolution allows UE to quickly detect SS by limiting the search over a limited number of center frequencies compared to search over carrier frequencies with channel raster of 15 kHz resolution. 

    


         \begin{figure*}[!b]
\centering
\includegraphics[width=1.85\columnwidth]{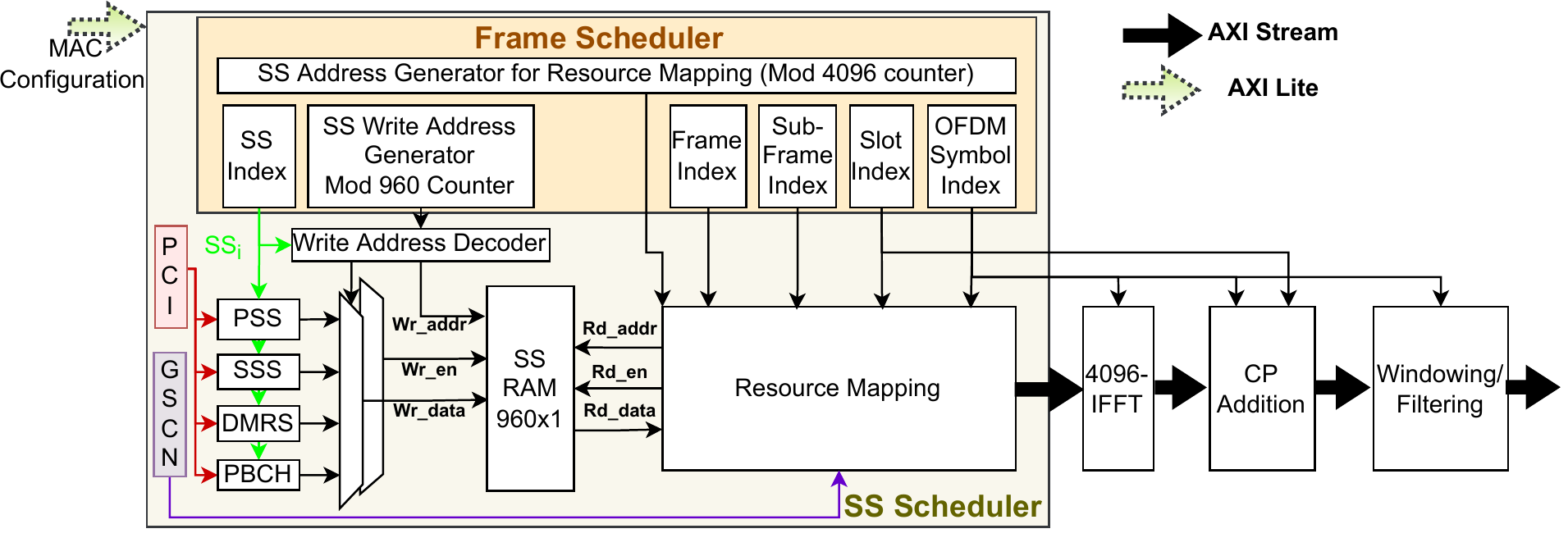}
\caption{Block diagram of CS transmitter PHY.}
\label{Fig:DL_PHY}
\vspace{-0.2cm}
\end{figure*}

\subsection{Review: Mapping of 5G CS PHY on Hardware}
The design, theoretical analysis, and simulation-based performance evaluation of 5G PHY have been an active research topic in the last few years \cite{SS3,SS2,SS1,PRACH1,PRACH2,5GPHY1,5GPHY2}. Various works on the design and optimization PHY sub-blocks such as channel encoders \cite{5GLDPC1,5GLDPC2,5GLDPC3,5GLDPC4}, OFDM modulator \cite{5GOFDM1,5GOFDM2}, beamforming \cite{5GBF1,5GBF2} and channel estimation \cite{5GCE1,5GCE2,5GCE3} have been explored to improve the PHY performance compared to 4G PHY. However, only a few works have focused on 3GPP standards and further improvements without compromising the compatibility with existing and previous standards \cite{LTE_SS,5GPHY2,SS3,SS1,PRACH2,PRACH1}. From CS perspectives, works in \cite{SS1,SS2} provide in-depth understanding of the SS generation while \cite{SS3} offers innovative intelligence based CS approach. However, these works do not highlight the challenges in SS detection and applications of SS detection for timing synchronization. The work presented in this paper aims to bridge this gap.  

Another important aspect of 5G PHY deployment is an efficient realization on the software and hardware platforms. Compared to 4G, there are numerous scenarios that have resulted in the split of PHY depending on positions of radio, distribution and centralized units \cite{5Gsplit1,5Gsplit2,5Gsplit3}. This 
demands hardware-software co-design of the PHY. In \cite{ofdm2,ofdm3}, authors have explored the hardware-software co-design of IEEE 802.11 PHY on Zynq system-on-chip (SoC) platform. In \cite{OFDM4,5GOFDM1}, authors have explored reconfigurable OFDM waveform with dynamically controlled out-of-band emission. In \cite{ofdm1}, authors have proposed reconfigurable OFDM-based PHY to support multi-standard operations. However, none of these works consider 3GPP compatible 5G PHY. The work in \cite{OFDM5} is limited to OFDM-based transceivers but does not consider various control and data channels in 5G. In this work, we design software and hardware IP cores of 3GPP compatible 5G PHY and demonstrate the CS operations by developing the receiver PHY. 
        
        \setcounter{figure}{4}
\begin{figure*}[!b]
\vspace{-0.2cm}
\centering
\includegraphics[width=1.9\columnwidth]{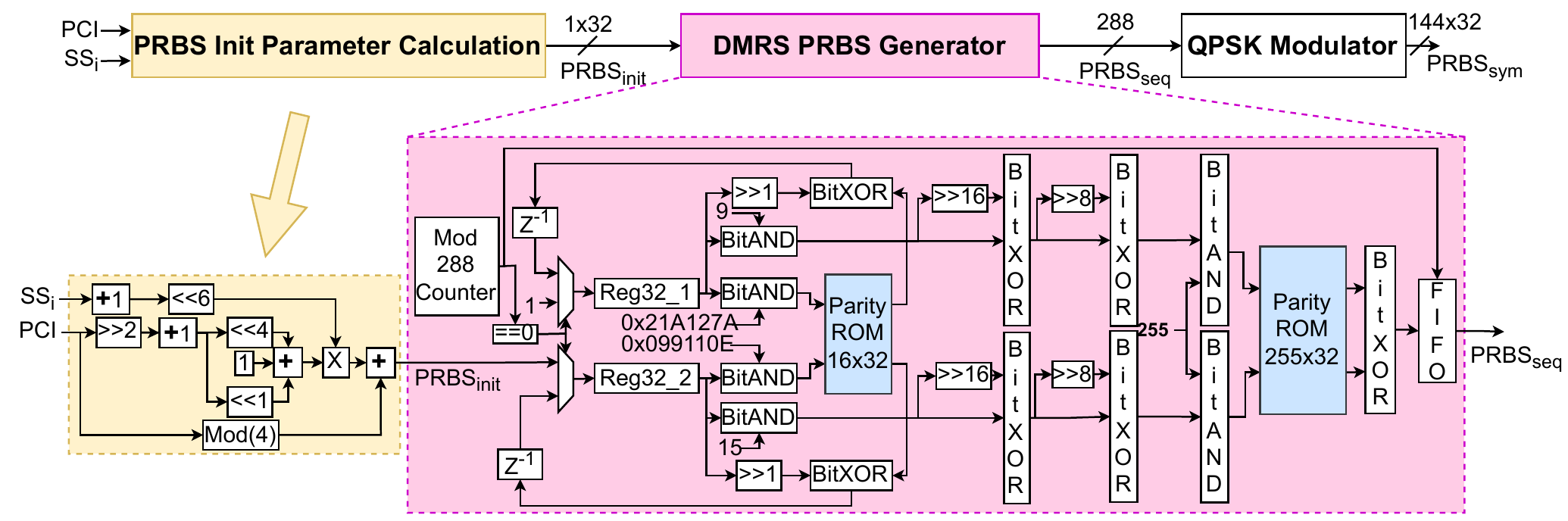}
\caption{Architecture for generating the 3GPP PBCH-DMRS.}
\label{fig:dmrsarch}
\end{figure*}

\section{Downlink Transmitter PHY for CS}
\label{Sec:DLphy}
In this section, we present the design details and architecture of the downlink transmitter PHY for CS. As shown in Fig.~\ref{Fig:DL_PHY}, the PHY consists of multiple scheduling and baseband signal processing tasks such as frame scheduler, SS scheduler, SS generation, 4096-IFFT, CP addition, and windowing or filtering. Among them, the last three blocks perform convectional OFDM modulation and are designed using FFT and filter IPs shared publicly by AMD-Xilinx. As discussed in Section~\ref{Sec:review}, the CP size depends on the location of the OFDM symbol in a slot, and hence, slot and OFDM symbol indexes are given as input to the CP addition block. 

SS scheduler is the most important unit of the CS PHY and it is responsible for generating the stream of OFDM symbols with embedded SS signals at the desired time and frequency locations as shown in Fig.~\ref{fig:SSB}. The SS scheduler mainly consists of a frame scheduler, generation blocks for PSS, SSS, DMRS, and PBCH, memory to store one instance of SS, and a resource mapper. The output of the SS scheduler is modulated using OFDM modulation. The 5G frame scheduler block is designed to generate multiple signals to keep the track of frame, sub-frame, slot, and OFDM boundaries. This is done using a modulus-4096 counter which is configured by the MAC layer. At the end of each cycle of this counter, the OFDM symbol index is incremented by 1. OFDM symbol index is tracked using the Modulus-14 counter and at the end of its cycle, the slot index is incremented by 1. In this way, slot index, sub-frame index, and frame index are generated using modulus-2, modulus-10, and modulus-1024 counters, respectively. The frame scheduler also generates the SS index, $SS_i$, to generate appropriate SS signals and keep the track of active SS transmission. The $SS_i$ is calculated using the OFDM symbol and frame indices as discussed in Section~\ref{Sec:review}. The final task of the frame scheduler is to generate the address generator so that 960 symbols corresponding to an upcoming instance of SS are stored in the memory. It is designed using a modulus-960 counter as shown in Fig.~\ref{Fig:DL_PHY}. Though each counter is shown as independent unit in Fig.~\ref{Fig:DL_PHY}, all counters are carefully implemented to maximize the shared hardware reuse and counters are synchronized so that the signals are generated as per the 3GPP timing requirements. 

For SS generation, PCI and GSCN values are configured by the MAC layer in the internal register via the AXI-Lite interface. The first sub-block is the PSS generator which outputs a 127-length binary phase shift key (BPSK) modulated sequence for a given PCI. It is based on the 3GPP specification in \cite{3GPP_211} 

\begin{equation}
    \text{PSS}(n) = 1 - 2 \text{PSS}_{ref}((n+43N_{ID}^{Cell}\%3)\%127)  \quad n \in \{0,1,..,126\}
\end{equation}
where 
\begin{equation*}
    \text{PSS}_{ref}(i+7) = [\text{PSS}_{ref}(i+4) + \text{PSS}_{ref}(i)] \%2 \quad i \in \{0,1,..,126\}
\end{equation*}

Here,  $\text{PSS}_{ref}(6:0)$ = \{1,1,1,0,1,1,0\}. We can see that the PSS sequence is obtained by reading the reference PSS sequence, $PSS_{ref}$ in a particular order, and this order is based on the input PCI, $N_{ID}^{Cell}$. Also, the modulus by 3 operations indicates that there are three candidate PSS sequences.

\setcounter{figure}{2}
\begin{figure}[!t]
\vspace{-0.2cm}
\centering
\includegraphics[width=0.9\columnwidth]{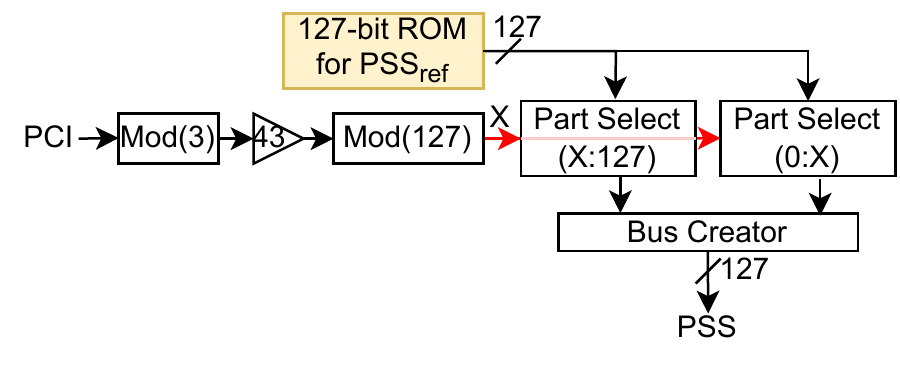}
\caption{Architecture for generating the 3GPP PSS.}
\label{fig:PSSarch}
\end{figure}

\begin{figure}[!t]
\centering
\vspace{-0.2cm}
\includegraphics[width=0.9\columnwidth]{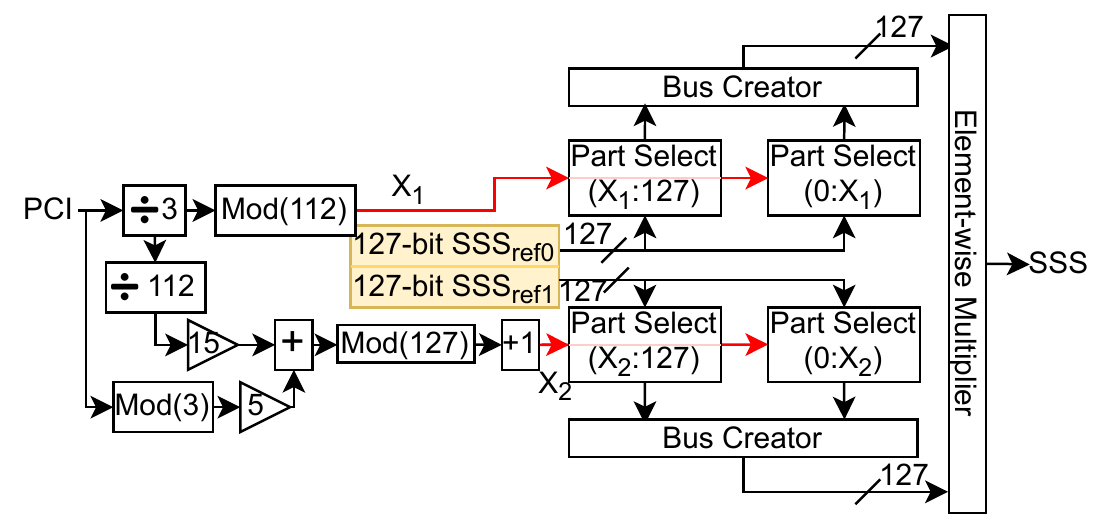}
\caption{Architecture for generating the 3GPP SSS.}
\label{fig:SSSarch}
\end{figure}

Similarly, SSS is also a 127-length BPSK modulated sequence, and one of the 336 candidate sequences is chosen using the input PCI, $N_{ID}^{Cell}$, as shown in Fig.~\ref{fig:SSSarch}. In SSS generation, two reference 127-lengths sequences are stored in the memory and they are read in a particular order followed by element-wise multiplication. Mathematically \cite{3GPP_211},

\begin{equation}
    \text{SSS}(n) = [1-2\text{SSS}_{ref0}(m_0)][1-2\text{SSS}_{ref1}(m_1)]  \quad n \in \{0,1,..,126\}
\end{equation}
where
\begin{equation}
    m_0 = \Bigg[n+\Bigg(15\Bigg\lfloor \frac{N_{ID}^{Cell}}{336} \Bigg\rfloor+5N_{ID}^{Cell}\%3\Bigg)\Bigg]\%127
\end{equation}
\begin{equation}
    m_1 = \Bigg[n+\Bigg(\frac{N_{ID}^{Cell}}{3}\Bigg) \% 112\Bigg]\% 127
\end{equation}
\begin{equation*}
    \text{SSS}_{ref0}(i+7) = [\text{SSS}_{ref0}(i+4) + \text{SSS}_{ref0}(i)] \%2 \quad i \in \{0,1,..,126\}
\end{equation*}
\begin{equation*}
    \text{SSS}_{ref1}(i+7) = [\text{SSS}_{ref1}(i+1) + \text{SSS}_{ref1}(i)] \%2 \quad i \in \{0,1,..,126\}
\end{equation*}

Here,  $\text{SSS}_{ref0}(6:0)$ = $\text{SSS}_{ref0}(6:0)$ = \{0,0,0,0,0,0,1\}. 

\setcounter{figure}{5}

         \begin{figure*}[!t]
\centering
\includegraphics[width=1.8\columnwidth]{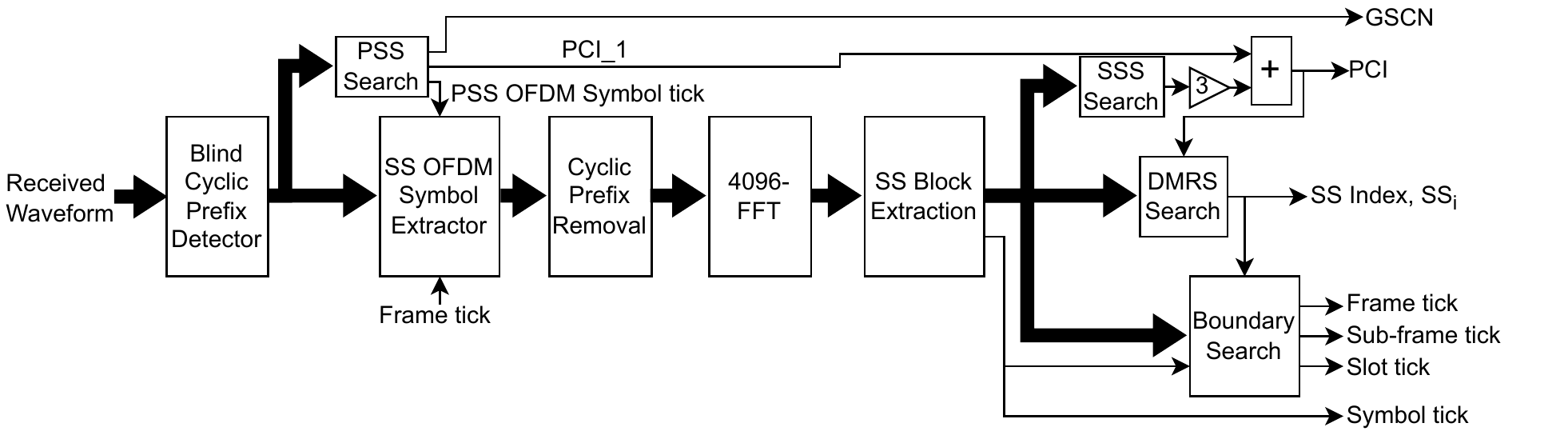}
\caption{Block diagram of the CS receiver PHY.}
\label{Fig:UL_PHY}
\end{figure*}

The DMRS is 144-length Quadrature Phase Shift Keying (QPSK) modulated sequence generated using the PCI, $N_{ID}^{Cell}$ and SS index, $SS_i$ as shown in Fig.~\ref{fig:dmrsarch}. Due to dependence on the value of $SS_i$, the DMRS is different for each SS in a SS burst. The DMRS generation involves pseudo-random binary sequence (PRBS) generation of length 288 which is then QPSK modulated to obtain the final DMRS. The PRBS generation is based on length-31 Gold sequence and it needs the initial seed, $PRBS_{init}$, which is calculated using the PCI, $N_{ID}^{Cell}$ and SS index, $SS_i$ \cite{3GPP_211}. 
\begin{equation}
        PRBS_{init} = 2^{11}(SS_i+1)\Bigg(\Bigg\lfloor\frac{N_{ID}^{Cell}}{4}\Bigg\rfloor+1\Bigg)+2^6(SS_i+1)+(N_{ID}^{Cell}\%4)
\end{equation}

Using the $PRBS_{init}$, $PRBS_{seq}$ of length 288 is generated using length-31 gold sequences \cite{3GPP_211}. The detailed steps and algorithm for PRBS generation have been discussed in \cite{3GPP_211} and the corresponding proposed hardware architecture is given in Fig.~\ref{fig:dmrsarch}. The final step in SS generation is the PBCH. Since it is not needed for the first phase of CS, we have randomly generated 432 complex symbols that are placed on the resources allocated for PBCH.


The generated SS block, i.e. four OFDM symbols each of 240 complex symbols, is then stored in the memory. The resource mapper block generates the stream of OFDM symbols of size 4096 SCs with an embedded SS at the desired time and frequency location. The frequency location is controlled by the GSCN configured by the MAC layer while the time location is as per the 3GPP specifications discussed in Section~\ref{Sec:review}: Fig.~\ref{fig:SSB}. Next, the design of the downlink receiver PHY is presented.

\section{Downlink Receiver PHY for CS}
\label{Sec:ULphy}
The downlink receiver PHY for CS needs to blindly detect the SS and identify the transmitted PCI, $N_{ID}^{Cell}$, and the index of the detected SS block, $SS_i$. It also helps the UE to locate the starting sample of each frame, sub-frame, slot, and symbol which is critical for receiving the MSI and establishing the uplink communication with the gNB.

The block diagram of the receiver PHY is shown in Fig.~\ref{Fig:UL_PHY}. The first step is to identify the presence of CP so that receiver can group the received samples in a packet of 4096 samples comprising a single OFDM symbol. This is done via a conventional auto-correlation-based approach where the received signal is auto-correlated with its delayed version since CP is the initial section of the OFDM symbol appended at the end. The CP detection enables the identification of symbol boundary so that 4096 complex symbols corresponding to a single OFDM symbol are extracted for the PSS search task. Since CP detection is a well-known task in wireless PHY, we skip the discussion on its architecture. Note that the CP detection is an optional step in CS PHY but it helps to reduce the number of samples to be processed by the PSS search block thereby speeding up the PSS search. The CP detection may fail when SNR is poor and in such cases, all the samples are forwarded to the PSS search block.

The next step is to identify the OFDM symbols containing the SS block. Though SS is transmitted in pre-defined OFDM symbols, the receiver is not aware of the index of the detected OFDM symbol and hence, blind detection of SS is needed.
We first detect the presence of the PSS using the PSS Search block and identify the detected PSS sequence, PCI\_2 $\in \{0,1,2\}$. The PSS detection also enables the extraction of symbol boundaries whenever CP detection fails.  
The OFDM symbol containing PSS can also be detected using the $frame\_tick$ signal generated by the boundary search block. However, in the beginning, $frame\_tick$ signal may not be available and hence, accurate and fast detection of PSS is critical for CS PHY. The SS OFDM Symbol Extractor block extracts the four OFDM symbols containing one instance of SS and forwards them to the OFDM demodulator containing the CP removal and 4096-FFT. After FFT, 960 symbols of the detected SS instance are extracted. Among them, samples belonging to SSS in the third OFDM symbol are sent to the SSS search block to identify the SSS sequence, PCI\_1 $\in \{0,1,..335\}$. Using the PSS and SSS search, we can estimate the PCI as 3*PCI\_1 + PCI\_2. The extracted 144 DMRS symbols are forwarded to the DMRS search unit to identify the SS index, $SS_i \in \{1,2,..,8\}$. In the end, the boundary search block exploits the timing information of the multiple SS blocks to identify the frame, sub-frame, and slot boundaries. Next, we present the architecture of the PSS search, SSS search, DMRS search, and boundary search blocks of the receiver CS PHY.

\subsection{PSS Search}
The PSS Search is the most important part of CS PHY as it helps the receiver to identify the presence of the SS block. The challenges in the PSS search are unknown symbol boundary, multiple locations since GSCN is unknown, and large transmission bandwidth of 100 MHz out of which SS occupies only 7.2 MHz resulting in large size correlators. 

\begin{figure}[!h]
\centering
\includegraphics[width=\columnwidth]{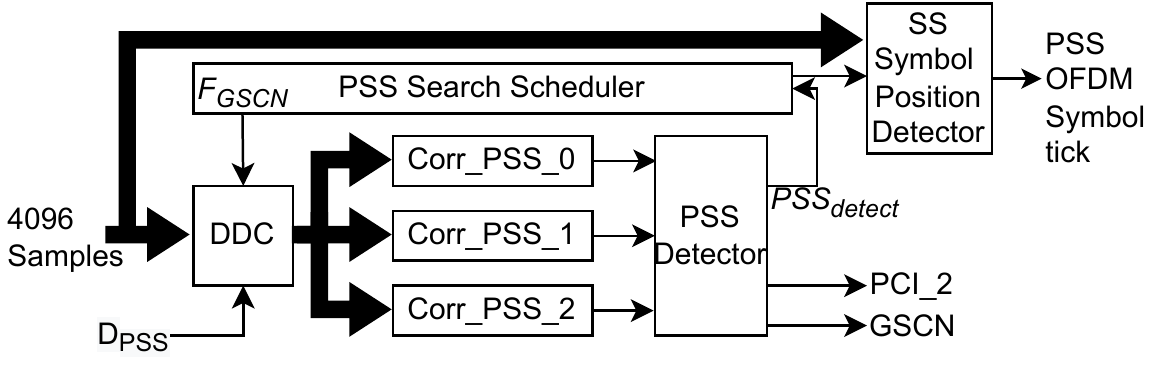}
\caption{PSS Search Architecture.}
\label{Fig: PSSSearch}
\end{figure}

To detect the transmitted PSS index, $PCI_1\in \{0,1,2\}$, the cross-correlation is performed between the three reference PSS signals and the received signal. Since the location or GSCN of the received PSS is unknown and SS bandwidth is small compared to total transmission bandwidth, direct cross-correlation between the received signal and reference PSS is computationally complex and time-consuming. Furthermore, generating separate PSS reference signals for each of the 340 different GSCN needs huge on-chip memory. To address these challenges, we propose a novel architecture for PSS search as shown in Fig.~\ref{Fig: PSSSearch}. In the proposed architecture, instead of generating reference PSS for every GSCN, we generate three reference PSS with the lowest GSCN, i.e., 7711. Thus, the received signal need to be down-converted to the lowest GSCN so that cross-correlation can be performed to detect the presence of the PSS. Since the GSCN of the received signal is not known, we need to perform down-conversion for all possible GSCN values and hence, a scheduler is used to generate appropriate down-conversion frequency, $f_{GSCN}$, sequentially till the PSS is detected. Since PSS occupies only 7.2 MHz bandwidth, the size of correlators can be reduced by downsampling reference as well as received signals. We have achieved these operations via a digital down converter (DDC) which downconverts the received signal to the GSCN value selected by the scheduler followed by the low-pass filtering to remove aliasing and down-sampling by the chosen factor, $D_{PSS}$.  Since the downsampling results in loss of signal and may affect the detection probability of the PSS, it should be chosen carefully based on the received signal-to-noise ratio (SNR).  Empirically, higher SNR allows the use of larger $D_{PSS}$ which in turn leads to faster PSS search and lower memory requirement. The final task after the PSS detection is to locate the starting sample of the OFDM symbol containing the PSS and generate an appropriate tick signal for further synchronization. This is done using the location of the correlation peak for the detected PSS and corresponding GSCN value. 


\subsection{SSS Search}
\begin{figure}[!h]
\centering
\includegraphics[width=\columnwidth]{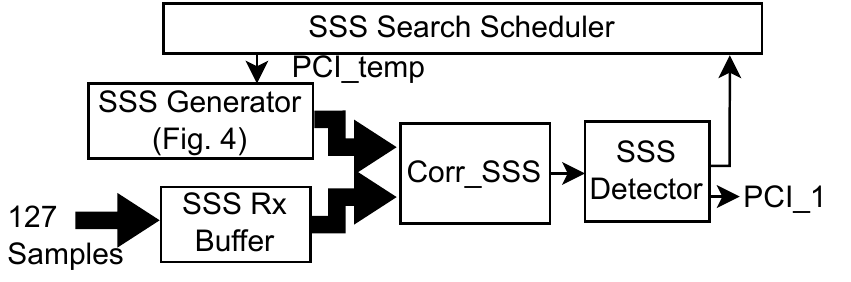}
\caption{SSS Search Architecture.}
\label{Fig:SSSSearch}
\end{figure}

In SSS Search, the received 240 samples extracted from the third OFDM symbol of the detected SS are correlated with 336 reference SSS sequences to detect the $PCI_2\in \{0,1,..,335\}$. Instead of storing all reference sequences, we have used the SSS generator in Fig.~\ref{fig:SSSarch} to generate the desired sequence. Since PCI is not known, the scheduler is used to cycle through all possible SSS sequences. The sequential SSS search architecture is shown in Fig.~\ref{Fig:SSSSearch}. Depending on the given resource and latency constraints, we can explore serial-parallel architecture by using multiple SSS generators and correlation blocks in parallel. 

\subsection{DMRS Search}
\begin{figure}[!h]
\centering
\includegraphics[width=\columnwidth]{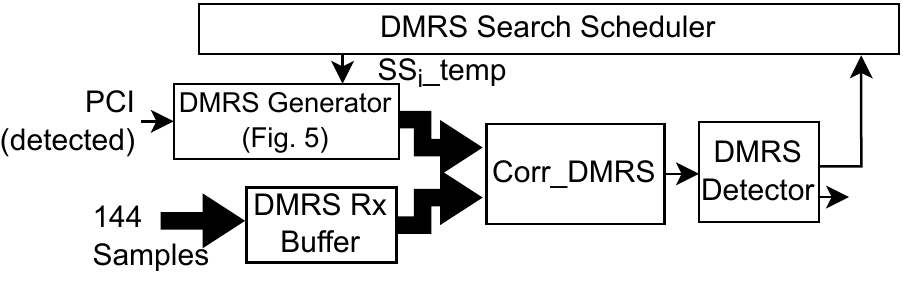}
\caption{DMRS Search Architecture.}
\label{Fig:DMRSSearch}
\end{figure}

The DMRS search block identifies the $SS_i \in \{0,1,..,7\}$ by correlating the 144 samples of received DMRS with eight candidate reference DMRS generated using the PCI detected by PSS and SSS search blocks. There are two ways to realize the DMRS search architecture: 1) Use the DMRS generator in Fig.~\ref{fig:dmrsarch} to generate reference DMRS, and 2) Store all possible DMRS, i.e., 1008*8=8064 sequences each consisting of 144 QPSK modulated samples which demand 9.3 MegaBytes of on-chip memory. Due to memory constraints on ZSoC, we have selected the first option and the corresponding architecture is shown in Fig.~\ref{Fig:DMRSSearch}. Similar to the SSS search, we can explore serial-parallel architecture by using multiple DMRS generators and correlation blocks in parallel.

\subsection{Boundary Search}
The aim of this block is to generate various time reference tick signals to locate the starting samples of the OFDM symbol, slot, sub-frame, and frame. These are critical for MSI reception and uplink synchronization. Such signals are also needed in 5G repeaters to accurately extract the data samples for re-transmission.

\begin{figure}[!t]
\centering
\includegraphics[width=\columnwidth]{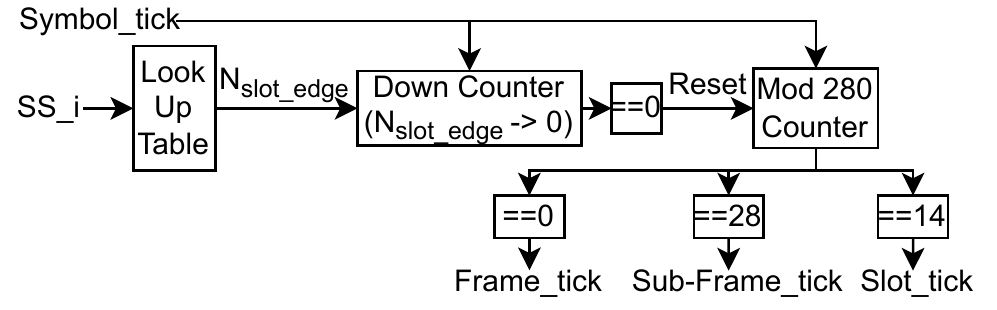}
\caption{Boundary Search Architecture.}
\label{Fig:boundarySearch}
\end{figure}

Since the SS signals are transmitted at fixed OFDM symbols and the DMRS search block has already identified the index of the detected SS signal, we can directly infer the OFDM symbol number for a detected \textit{Symbol\_tick} by PSS search and SS Block extraction blocks. Since each slot, sub-frame, and frame consists of a certain number of OFDM symbols, corresponding ticks are appropriately generated as shown in Fig.~\ref{Fig:boundarySearch}. For instance, detection of $SS_i=2$ corresponds to OFDM symbol number of 8 for PSS which lies in slot 0 of the even frame. Using this information, \textit{Slot\_tick} is detected which goes high after $N_{slot\_edge}=6$ OFDM symbols from the detected PSS. Similarly, the detection of $SS_i=3$ corresponds to the OFDM symbol number of 16 for PSS which lies in slot 1 of the even frame. Using this information, \textit{Slot\_tick} is detected which goes high after $N_{slot\_edge}=12$ OFDM symbols from the detected PSS. Thereafter, \textit{Slot\_tick} is generated after the interval of every 14 OFDM symbols. For every two low-to-high transitions of the \textit{Slot\_tick}, \textit{Sub-frame\_tick} is generated and for every 10 low-to-high transitions of the \textit{Sub-frame\_tick}, \textit{Frame\_tick} is generated.

\section{Performance and Complexity Analysis}
\label{Sec:results}
In this section, we present the functional correctness and complexity analysis of the proposed architecture for different SNR, WL, and PSS down-sampling factors. We specifically focus on the PSS Search, SSS Search, and DMRS Search blocks using the data samples generated by the transmit PHY.

\subsection{PSS Search}
\begin{figure}[!t]
\centering
\includegraphics[width=\columnwidth]{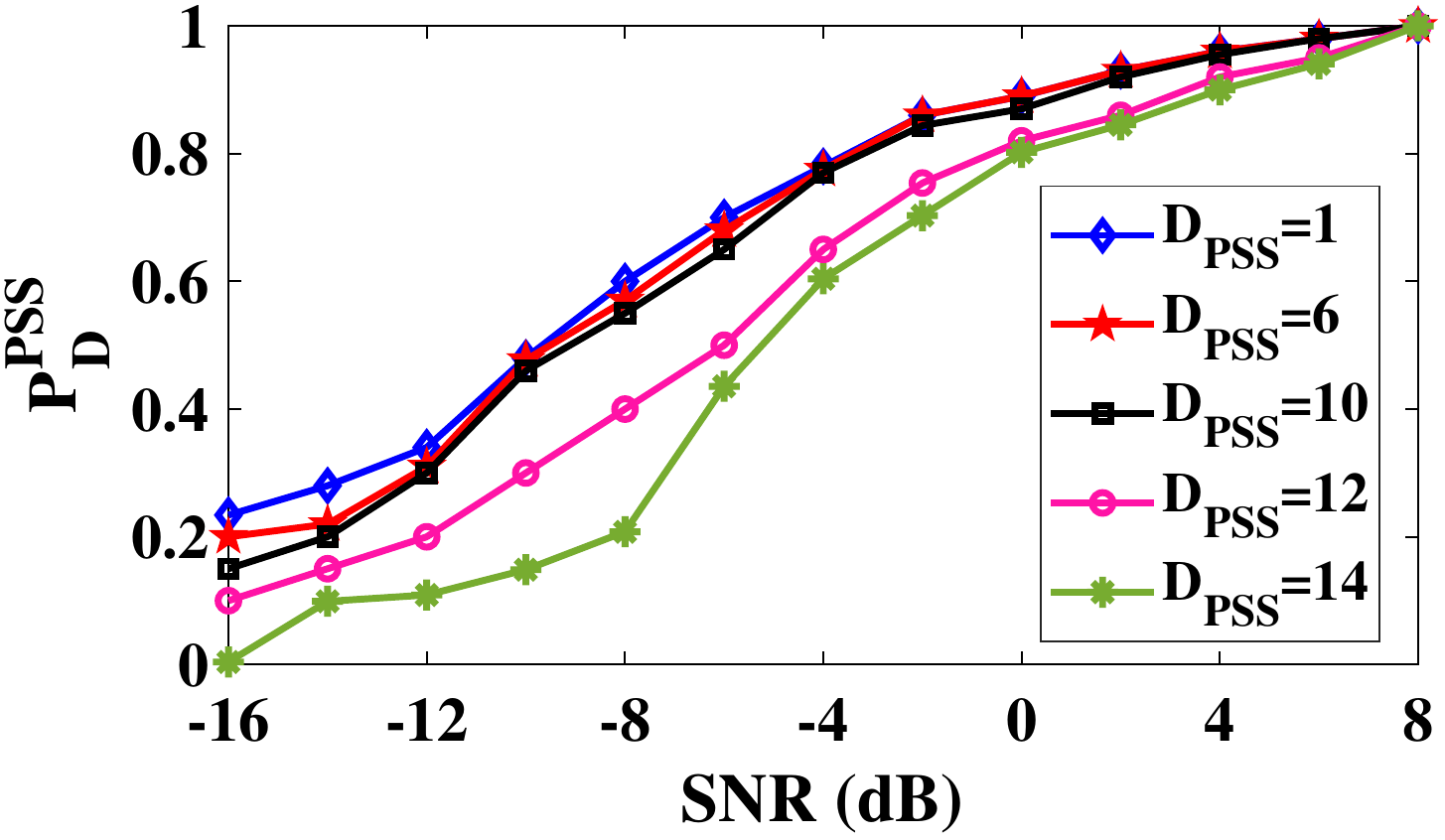}
\caption{Effect of SNR and $D_{PSS}$ on the probability of detection, $P_D^{PSS}$, for PSS search architecture.}
\label{Fig:PSS_det_dwns}
\end{figure}

The main task of the PSS Search is to identify the PCI\_2 and GSCN. Since the identification of  PCI\_2 also confirms the correct identification of the GSCN, we limit the discussion on the probability of detection of  PCI\_2 for different SNRs and $D_{PSS}=\{1,6,10,12,14\}$. It is referred to as $P_D^{PSS}$. In Fig.~\ref{Fig:PSS_det_dwns}, $P_D^{PSS}$ obtained from the experimental results on single-precision floating-point (SPFL) architecture for SNR ranging from -16 dB to 8 dB and $D_{PSS}=\{1,6,10,12,14\}$ are shown. It can be observed that $P_D^{PSS}$ increases with the increase in SNR and $D_{PSS}$. The degradation in the $P_D^{PSS}$ is not significant for $D_{PSS} \leq 10$ even at low SNR when compared to PSS Search without any downsampling, i.e., $D_{PSS}=1$. At SNR higher than 5 dB, it is possible to perform PSS Search with $D_{PSS}$ of 14 as well. Experimental results show that PSS Search fails for $D_{PSS} \geq 14$ due to the insufficient number of samples available for correlation with reference PSS.

\begin{figure}[!t]
\centering
\includegraphics[width=\columnwidth]{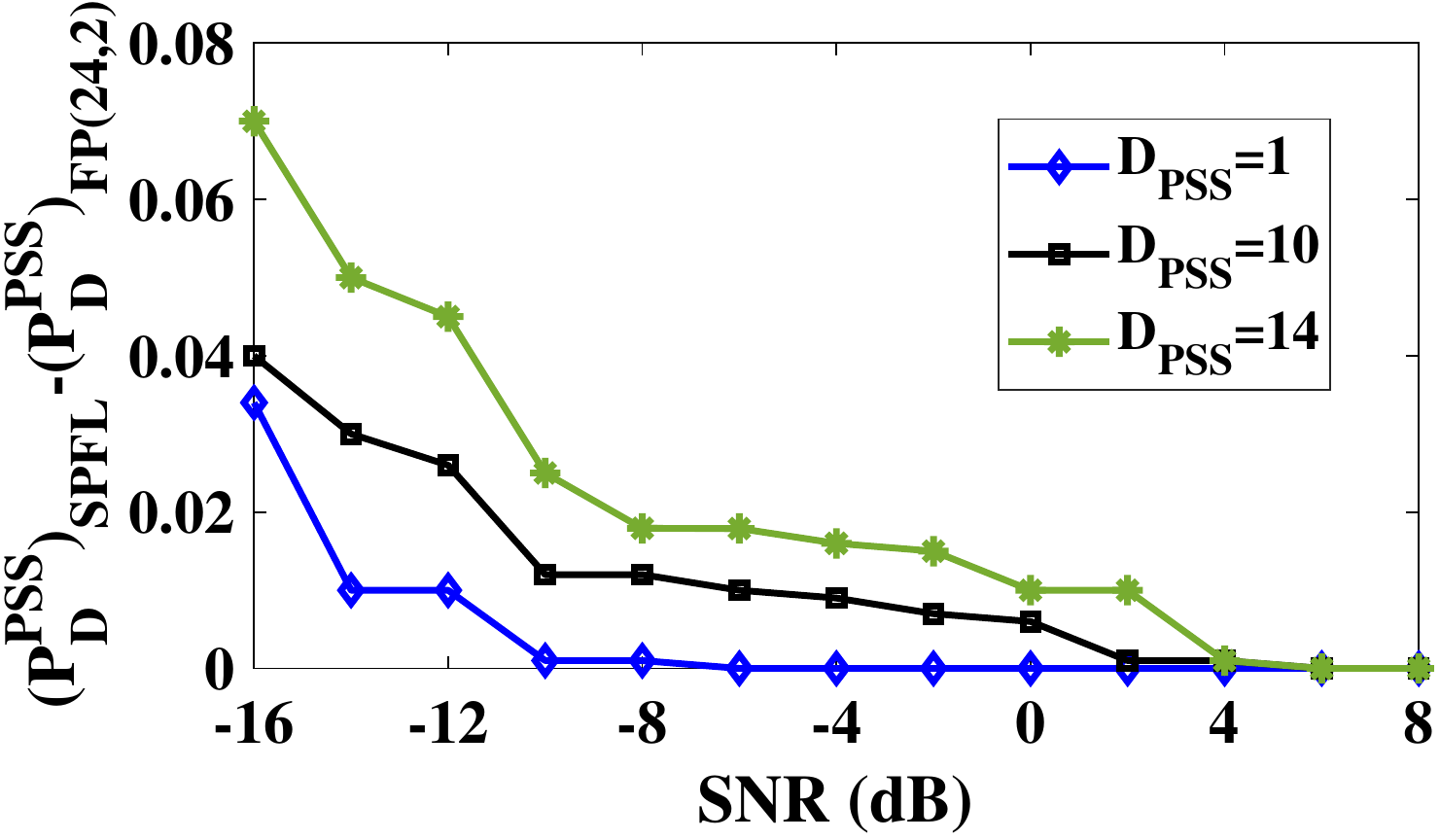}
\caption{Effect of WL on the probability of detection, $P_D^{PSS}$, for PSS search architecture.}
\label{Fig:PSS_det_dwns_WL}
\end{figure}

Next, we analyze the effect of WL of the PSS search architecture on the $P_D^{PSS}$. We consider two fixed-point architectures with a total WL of 32 and 24 bits. The WL of 32 bits is selected to keep the number of bits the same as that of SPFL architecture. Experimental results showed that the PSS detection fails for any WL lower than 24 bits and hence, second architecture with a WL of 24 bits is selected. In each fixed-point architecture, the number of bits allocated to integer and fractional parts is carefully selected to maximize the probability of detection. In Fig.~\ref{Fig:PSS_det_dwns_WL}, we compare the difference in $P_D^{PSS}$ for SPFL and fixed-point 24-bit architecture for SNR ranging from -16 dB to  8 dB and $D_{PSS}=\{1,10,14\}$. It can be observed that the error is small and negligible for $D_{PSS} \leq 10$ even at low SNR.

The architecture with lower WL and higher $D_{PSS}$  offers significant savings in resource utilization and execution time. In Table~\ref{Tab:PSS_Table}, we first compare the execution time in milliseconds (ms). It can be observed that the execution time decreases with the increase in the $D_{PSS}$. Also, fixed-point architectures offer a 2-3 factor reduction in the execution time over SPFL architecture. Similarly, higher $D_{PSS}$ leads to lower utilization of on-chip memory, i.e. FPGA block RAM, due to a few samples of the received and reference PSS. The use of lower WL also offers a further reduction in on-chip memory. Interestingly, the effect of $D_{PSS}$ on the number of embedded multipliers and FPGA LUTs is not significant since the correlation block is realized in a sequential manner due to the limited number of memory ports. However, lower WL offers further savings in multipliers and LUTs since the size of multipliers and accumulators is smaller for lower WL. To summarize, the appropriate selection of $D_{PSS}$ and WL is important to meet the desired execution time, functional accuracy, and resource utilization constraints. For such analysis, the proposed work of mapping of complete transmit and receiver PHY on the SoC is critical.

\begin{table}[!h]
\centering
\caption{Execution Time and Resource Utilization Comparison for PSS Search}
\label{Tab:PSS_Table}
\renewcommand{\arraystretch}{1.2}
\resizebox{\columnwidth}{!}{%
\begin{tabular}{|c|l|lll|}
\hline
\multirow{2}{*}{\textbf{Parameters}} & \multicolumn{1}{c|}{\multirow{2}{*}{\textbf{Word Length}}} & \multicolumn{3}{c|}{{$D_{PSS}$}} \\ \cline{3-5} 
 & \multicolumn{1}{c|}{} & \multicolumn{1}{c|}{\textbf{1}} & \multicolumn{1}{c|}{\textbf{6}} & \multicolumn{1}{c|}{\textbf{10}} \\ \hline
\multirow{3}{*}{\textbf{\begin{tabular}[c]{@{}c@{}}Execution Time \\ (ms)\end{tabular}}} & \textbf{SPFL (32 bits)} & \multicolumn{1}{l|}{29.3} & \multicolumn{1}{l|}{19.2} & 19.1 \\ \cline{2-5} 
 & \textbf{Fixed Point \{32,2\}} & \multicolumn{1}{l|}{\textbf{13.3}} & \multicolumn{1}{l|}{6.9} & 6.85 \\ \cline{2-5} 
 & \textbf{Fixed Point \{24,2\}} & \multicolumn{1}{l|}{\textbf{13.3}} & \multicolumn{1}{l|}{\textbf{6.8}} & \textbf{6.84} \\ \hline
\multirow{3}{*}{\textbf{\begin{tabular}[c]{@{}c@{}}On Chip Memory \\ (18 KB BRAMs)\end{tabular}}} & \textbf{SPFL (32 bits)} & \multicolumn{1}{l|}{156} & \multicolumn{1}{l|}{108} & 100 \\ \cline{2-5} 
 & \textbf{Fixed Point \{32,2\}} & \multicolumn{1}{l|}{144} & \multicolumn{1}{l|}{106} & 96 \\ \cline{2-5} 
 & \textbf{Fixed Point \{24,2\}} & \multicolumn{1}{l|}{\textbf{108}} & \multicolumn{1}{l|}{\textbf{82}} & \textbf{80} \\ \hline
\multirow{3}{*}{\textbf{\begin{tabular}[c]{@{}c@{}}Embedded \\Multipliers \\ (DSP48s)\end{tabular}}} & \textbf{SPFL (32 bits)} & \multicolumn{1}{l|}{137} & \multicolumn{1}{l|}{137} & 137 \\ \cline{2-5} 
 & \textbf{Fixed Point \{32,2\}} & \multicolumn{1}{l|}{107} & \multicolumn{1}{l|}{107} & 107 \\ \cline{2-5} 
 & \textbf{Fixed Point \{24,2\}} & \multicolumn{1}{l|}{\textbf{104}} & \multicolumn{1}{l|}{\textbf{104}} & \textbf{104} \\ \hline
\multirow{3}{*}{\textbf{\begin{tabular}[c]{@{}c@{}}6-input \\Look-Up-Table \\ (LUT)\end{tabular}}} & \textbf{SPFL (32 bits)} & \multicolumn{1}{l|}{42.7 K} & \multicolumn{1}{l|}{42.6 K} & 42.6 K \\ \cline{2-5} 
 & \textbf{Fixed Point \{32,2\}} & \multicolumn{1}{l|}{41.5 K} & \multicolumn{1}{l|}{41.3 K} & 41.3 K \\ \cline{2-5} 
 & \textbf{Fixed Point \{24,2\}} & \multicolumn{1}{l|}{\textbf{31.1 K}} & \multicolumn{1}{l|}{\textbf{30.9 K}} & \textbf{30.9 K} \\ \hline
\end{tabular}%
}
\end{table}

\subsection{SSS Search}
\begin{figure}[!b]
\centering
\includegraphics[width=\columnwidth]{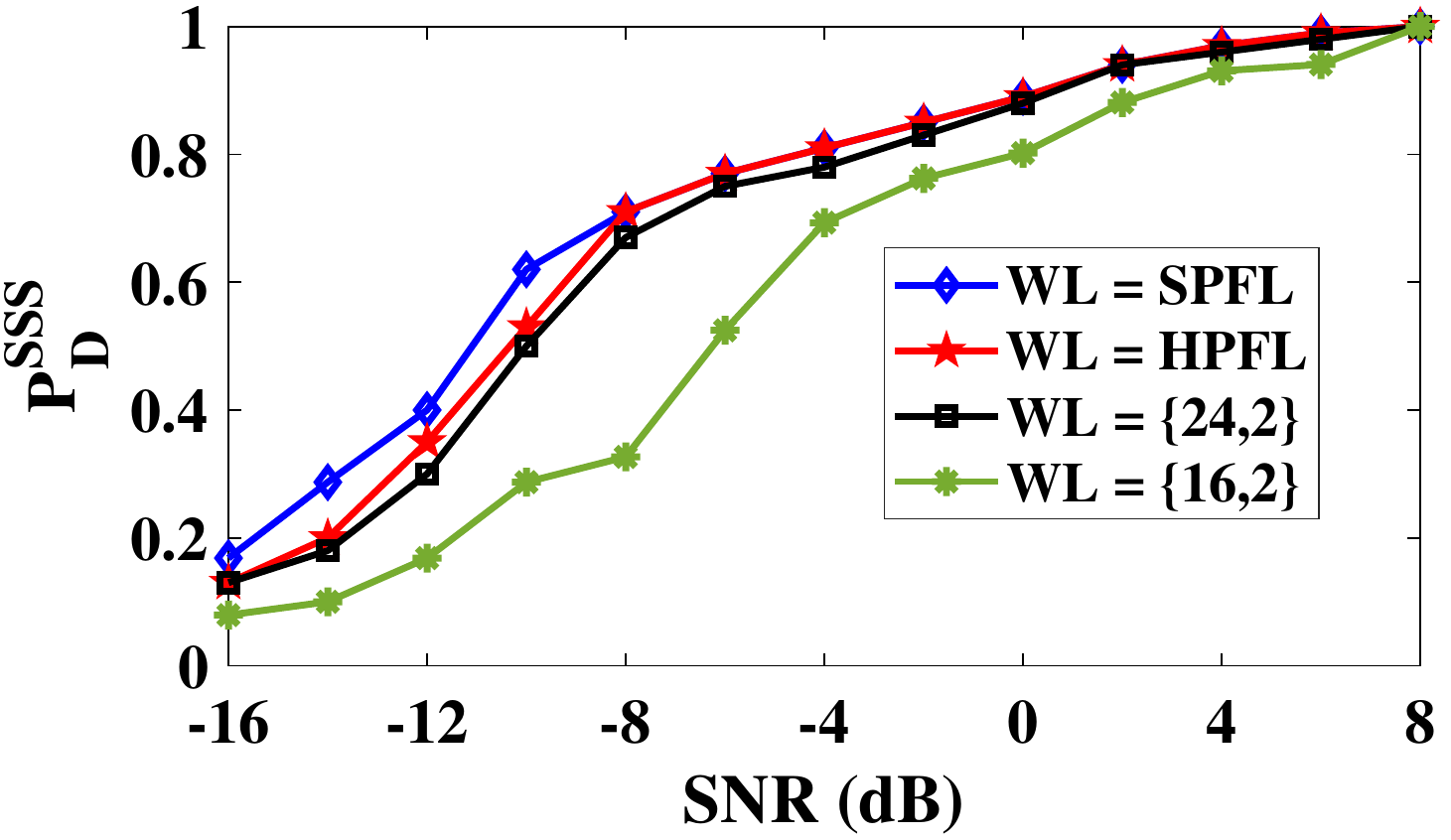}
\caption{Effect of WL on the probability of detection, $P_D^{SSS}$, for SSS search architecture.}
\label{Fig:SSS_det_dwns}
\end{figure}

The main task of the SSS Search is to identify the PCI\_1 which along with PCI\_2 from PSS Search gives PCI. The SSS Search is computationally less complex than PSS Search due to the small correlation size and hence, there is no need for down-sampling. We analyze the effect of WL on the probability of detection of PCI\_1, referred to as $P^{SSS}$, for different SNRs. We consider four different WL as shown in Fig.~\ref{Fig:SSS_det_dwns} and corresponding execution time and resource utilization results are given in Table~\ref{Tab:SSS_Table}. It can be observed that SPFL, half-precision floating-point (HPFL) and fixed point architecture with 24 bits offer nearly identical detection performance. On the other hand, execution time decreases by half as we move from SPFL to fixed-point architecture. In addition, there are significant savings in resource utilization as well. Further savings in resource utilization is possible using a fixed-point architecture with WL of 16 bits but it incurs slight degradation in detection performance. When compared to PSS search, SSS search is significantly faster and requires lower on-chip memory due to a smaller correlation size. The higher number of DSP48s and LUTs are due to serial-parallel realization of correlators as a complete SSS search involves 336 correlation operations compared to 3 in the PSS search.

\begin{table}[!h]
\centering
\caption{Execution Time and Resource Utilization Comparison for SSS Search}
\label{Tab:SSS_Table}
\renewcommand{\arraystretch}{1.2}
\resizebox{\columnwidth}{!}{%
\begin{tabular}{|c|c|c|c|c|}
\hline
\textbf{Parameters} & \textbf{SPFL} & \textbf{HPFL} & \textbf{\begin{tabular}[c]{@{}c@{}}Fixed Point \\ \{24,2\}\end{tabular}} & \textbf{\begin{tabular}[c]{@{}c@{}}Fixed Point \\ \{16,2\}\end{tabular}} \\ \hline
\textbf{Execution Time ($\mu$s)} & 33.4 & 18.8 & 16.6 & 16.6 \\ \hline
\textbf{\begin{tabular}[c]{@{}c@{}}On Chip Memory \\ (18 KiloBytes BRAMs)\end{tabular}} & 58 & 49 & 44 & 31 \\ \hline
\textbf{\begin{tabular}[c]{@{}c@{}}Embedded Multipliers \\ (DSP48s)\end{tabular}} & 651 & 573 & 573 & 558 \\ \hline
\textbf{\begin{tabular}[c]{@{}c@{}}6-input Look-Up-Table \\ (LUT)\end{tabular}} & 84052 & 33696 & 32574 & 28878 \\ \hline
\end{tabular}%
}
\end{table}

\subsection{DMRS Search}
\begin{figure}[!b]
\centering
\includegraphics[width=\columnwidth]{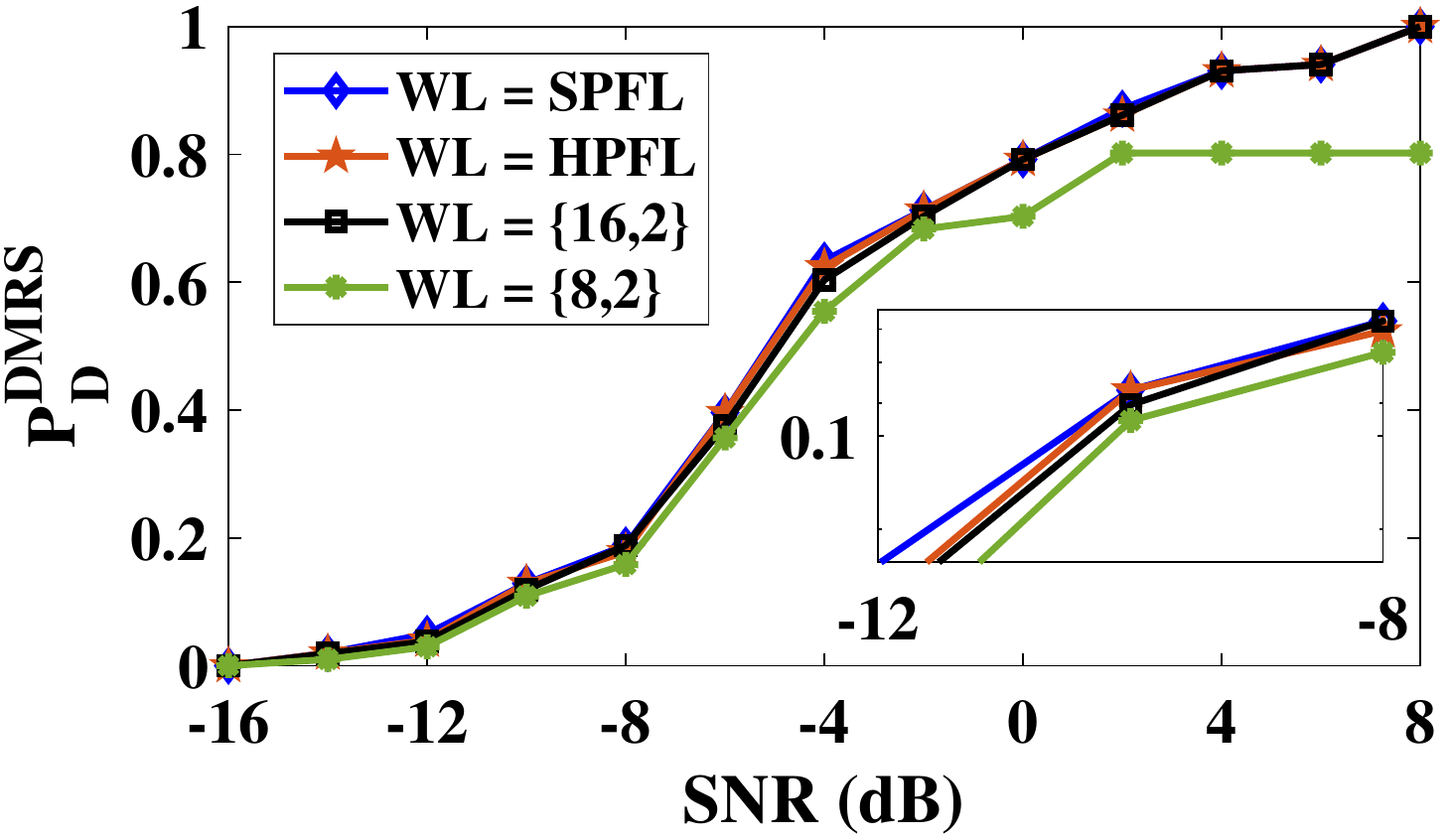}
\caption{Effect of WL on the probability of detection, $P_D^{SSS}$, for DMRS search architecture.}
\label{Fig:dmrsresults}
\end{figure}

Similar to PSS and SSS search, we compare the probability of detection of SS index, referred to as $P^{DMRS}_D$, for different WL and SNRs. As shown in Fig.~\ref{Fig:dmrsresults}, we can observe that WL of 16 offers nearly identical performance to that of SPFL and HPFL architecture. As shown in Table~\ref{Tab:DMRS_Table}, DMRS search architecture with WL of 16 offers 33\% reductions in execution time along with significant savings in resource utilization as well. 

For all the results presented in this section, we have used the transmitter PHY with a WL of 16. This is done by observing the power spectral density of the transmitter output and performance of PSS search with SPFL WL for different WLs of transmitter PHY. Corresponding results are not included here to avoid repetition of results.

\begin{table}[!b]
\centering
\caption{Execution Time and Resource Utilization Comparison for DMRS Search}
\label{Tab:DMRS_Table}
\renewcommand{\arraystretch}{1.2}
\resizebox{\columnwidth}{!}{%
\begin{tabular}{|c|c|c|c|c|}
\hline
\textbf{Parameters} & \textbf{SPFL} & \textbf{HPFL} & \textbf{\begin{tabular}[c]{@{}c@{}}Fixed Point \\ \{16,2\}\end{tabular}} & \textbf{\begin{tabular}[c]{@{}c@{}}Fixed Point \\ \{8,2\}\end{tabular}} \\ \hline
\textbf{Execution Time ($\mu$s)} & 6.7 & 6.1 & 2.1 & 1.9 \\ \hline
\textbf{\begin{tabular}[c]{@{}c@{}}On Chip Memory \\ (18 KiloBytes BRAMs)\end{tabular}} & 83 & 60 & 60 & 42 \\ \hline
\textbf{\begin{tabular}[c]{@{}c@{}}Embedded Multipliers \\ (DSP48s)\end{tabular}} & 85 & 71 & 70 & 66 \\ \hline
\textbf{\begin{tabular}[c]{@{}c@{}}6-input Look-Up-Table \\ (LUT)\end{tabular}} & 50311 & 39956 & 38933 & 28069 \\ \hline
\end{tabular}%
}
\end{table}

   \begin{figure*}[!t]
\centering
\includegraphics[width=1.9\columnwidth]{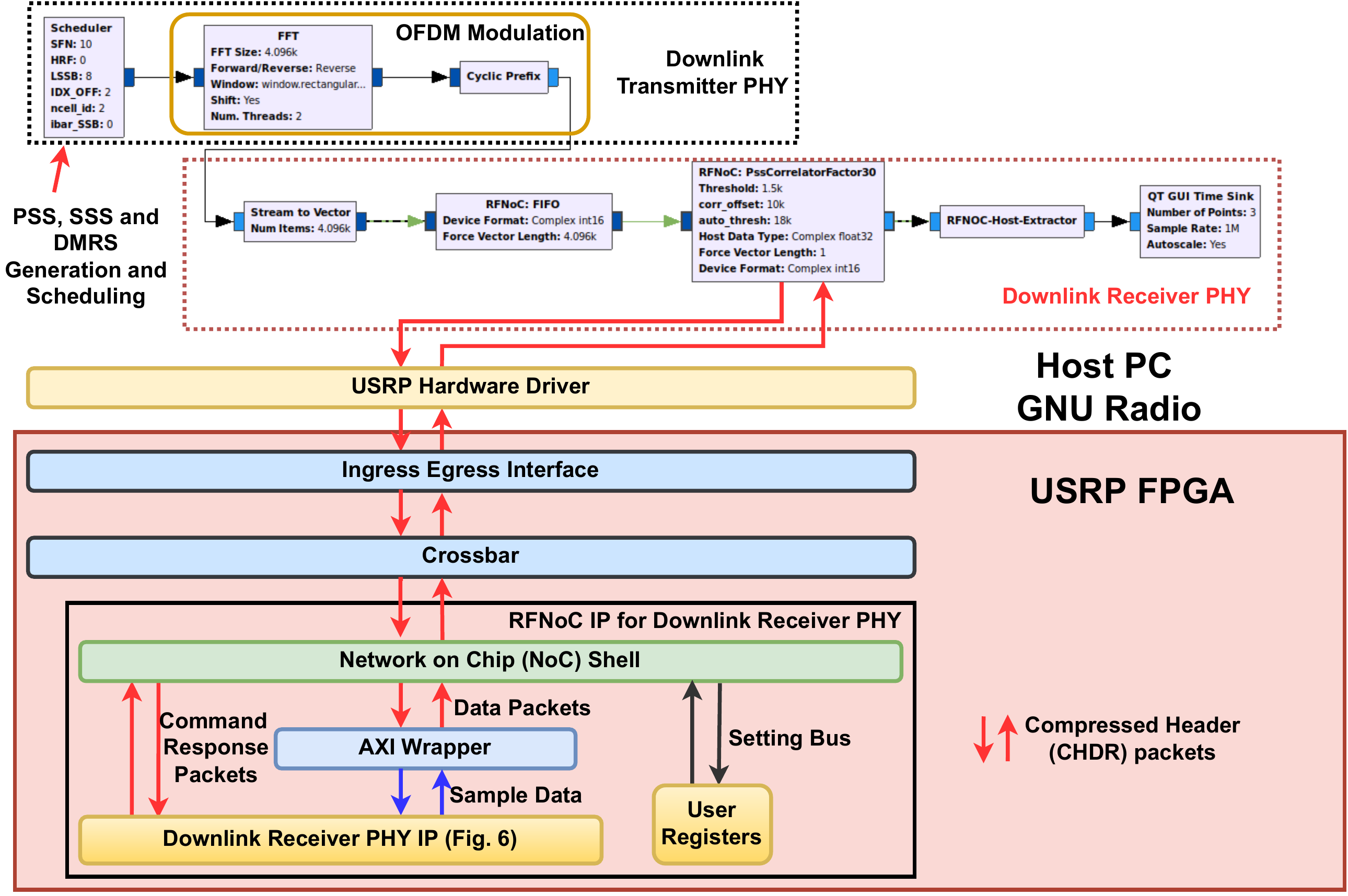}
\caption{Realization of the CS transmitter and receiver PHY using GNU Radio based RFNoC framework on X310 USRP platform.}
\label{Fig:rfnocarch}
\end{figure*}

\section{PHY Deployment on RFNoC Platform}
\label{Sec:RFNOC}
In academia as well as industry, GNU Radio is widely used for prototyping wireless systems and integration with the radio front-end of the USRP platform for demonstration in a real-radio environment. Since the GNU Radio tool is deployed on the host processor, the performance of signal processing algorithms is limited due to its sequential nature. To address this issue, Ettus Research developed the RFNoC tool that enables the acceleration of GNU Radio algorithms on hardware such as USRP FPGA \cite{RFNOC1,RFNOC2}. It offers a seamless tight interface between GNU Radio software and FPGA for data transfer thus enabling hardware-software co-design between a host processor and FPGA. Like GNU Radio, RFNoC is free and open-source software. 

In this work, we design complete software realization of the CS transmitter and receiver PHY using GNU Radio. Next, we have modified the AXI-Stream compatible hardware IPs discussed in Sections \ref{Sec:DLphy} and \ref{Sec:ULphy} for different building blocks of the CS transmitter and receiver PHY, respectively, into the custom RFNoC hardware IPs. Then, we have verified the functional correctness of these IPs using custom test benches and an out-of-tree module-based approach which allows integration and verification of RFNoC blocks in GNU Radio. For instance, one of the configurations with CS transmitter PHY in GNU Radio and CS receiver PHY on USRP FPGA is shown in Fig.~\ref{Fig:rfnocarch}. As per the requirement of the RFNoC framework, the proposed receiver PHY hardware IP with AXI-stream interface is integrated with the AXI wrapper and network-on-chip (NoC) shell. This integrated IP is capable of transmitting and receiving data packets from GNU Radio over an Ethernet interface using the RFNoC framework. The received packets from the transmitter are decoded by the NoC shell to extract the data samples which are then forwarded to receiver PHY along with appropriate control signals via the settings bus. The processed data is then sent back to GNU Radio via another packet using the NoC shell. In the same fashion, multiple blocks in GNU Radio can be moved to USRP FPGA to improve the execution time via parallel processing on the FPGA. 

In Table~\ref{tab:rfpss}, we compare the execution time of the PSS search block on GNU Radio and RFNoC platforms. We have selected SPFL WL for a fair comparison between both platforms. Similar to Table~\ref{Tab:PSS_Table}, the execution time decreases with the increase in $D_{PSS}$. It can be observed that the RFNoC-based acceleration offers an improvement in execution time by a factor 2 or higher. Similar to Table~\ref{Tab:PSS_Table}, higher $D_{PSS}$ and lower WL offer savings in resource utilization on the RFNoc platform as well, and corresponding results are skipped to avoid repetition of discussion. 

\begin{table}[!h]
\centering
\caption{Execution time of the PSS search IP in milliseconds (ms)}
\label{tab:rfpss}
\renewcommand{\arraystretch}{1.2}
\resizebox{0.8\columnwidth}{!}{%
\begin{tabular}{|l|cccc|}
\hline
\multirow{2}{*}{\textbf{Platform}} & \multicolumn{4}{c|}{\textbf{$D_{PSS}$}} \\ \cline{2-5} 
 & \multicolumn{1}{c|}{\textbf{1}} & \multicolumn{1}{c|}{\textbf{6}} & \multicolumn{1}{c|}{\textbf{10}} & \textbf{14} \\ \hline
\textbf{GNU Radio} & \multicolumn{1}{c|}{27.6} & \multicolumn{1}{c|}{17.9} & \multicolumn{1}{c|}{15.3} & 13.6 \\ \hline
\textbf{RFNoC} & \multicolumn{1}{c|}{9.54} & \multicolumn{1}{c|}{8} & \multicolumn{1}{c|}{7.2} & 6.5 \\ \hline
\end{tabular}%
}
\end{table}

\begin{figure}[!h]
\centering
\includegraphics[width=\columnwidth]{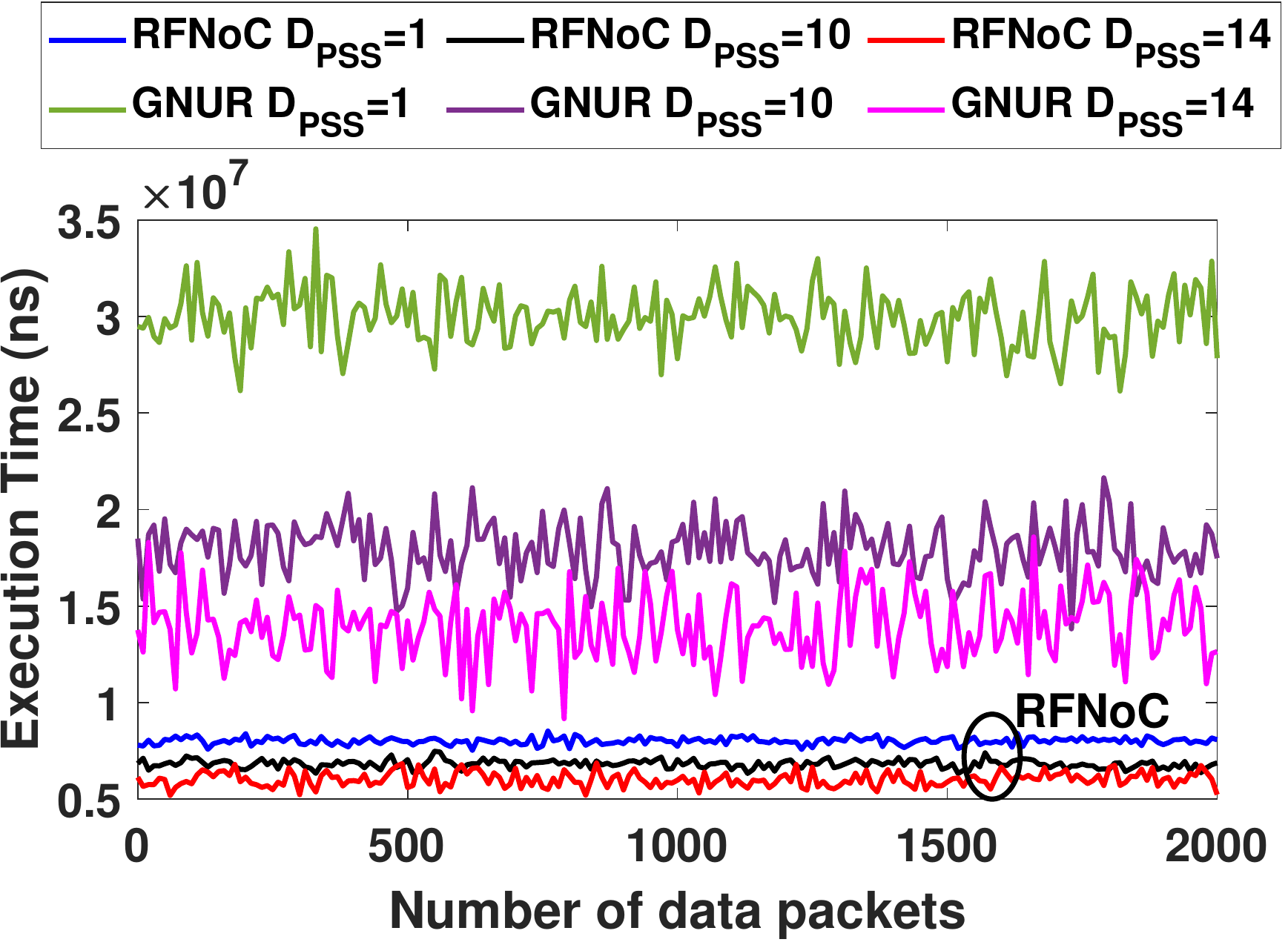}
\caption{Execution time comparison for PSS detection on GNU Radio and RFNoC for different downsampling factors. Here we consider single SS burst comprising of 8 SS signals over two frames.}
\label{fig:rfnoc}
\end{figure}

Next, we consider the end-to-end execution time which involves the data communication overhead between GNU Radio and USRP. We consider around two frames of data which corresponds to around 2000 data packets. As shown in Fig.~\ref{fig:rfnoc}, RFNoC based architecture outperforms the GNU Radio based architecture. In addition, variation in execution time of packets is lower in RFNoC compared to GNU Radio thereby offering stable and reliable performance. 

\section{Conclusions and Future Works}
\label{Sec:Conc}
In this work, we studied and designed hardware IP cores for 5G cell search (CS) as per the third generation partnership project (3GPP) specifications. Our contributions include the design of CS transmitter PHY consisting of synchronization signal (SS) generation, resource mapping, scheduler, and orthogonal frequency division multiplexing (OFDM) modulation and receiver PHY consisting of blind primary SS search, OFDM demodulation, secondary SS (SSS) search, a demodulation reference signal (DMRS) search and boundary detection for the frame, sub-frame, slot, and symbols. We have proposed a novel down-sampling approach for PSS search which offers a substantial reduction in execution time and resource utilization. We have demonstrated the functional correctness and superiority of the proposed approach via the RFNoC framework. Future works include an extension of the proposed architecture for physical broadcast channel (PBCH) detection, physical random access channel (PRACH) generation, and detection.

\bibliography{REF}

\begin{thebibliography}{10}
\providecommand{\url}[1]{#1}
\csname url@samestyle\endcsname
\providecommand{\newblock}{\relax}
\providecommand{\bibinfo}[2]{#2}
\providecommand{\BIBentrySTDinterwordspacing}{\spaceskip=0pt\relax}
\providecommand{\BIBentryALTinterwordstretchfactor}{4}
\providecommand{\BIBentryALTinterwordspacing}{\spaceskip=\fontdimen2\font plus
\BIBentryALTinterwordstretchfactor\fontdimen3\font minus
  \fontdimen4\font\relax}
\providecommand{\BIBforeignlanguage}[2]{{%
\expandafter\ifx\csname l@#1\endcsname\relax
\typeout{** WARNING: IEEEtran.bst: No hyphenation pattern has been}%
\typeout{** loaded for the language `#1'. Using the pattern for}%
\typeout{** the default language instead.}%
\else
\language=\csname l@#1\endcsname
\fi
#2}}
\providecommand{\BIBdecl}{\relax}
\BIBdecl

\bibitem{CSreview}
S.~Won and S.~W. Choi, ``Three decades of 3gpp target cell search through 3g,
  4g, and 5g,'' \emph{IEEE Access}, vol.~8, pp. 116\,914--116\,960, 2020.

\bibitem{IA1}
S.-Y. Lien, S.-L. Shieh, Y.~Huang, B.~Su, Y.-L. Hsu, and H.-Y. Wei, ``5g new
  radio: Waveform, frame structure, multiple access, and initial access,''
  \emph{IEEE Communications Magazine}, vol.~55, no.~6, pp. 64--71, 2017.

\bibitem{IA2}
E.~Dahlman and S.~Parkvall, ``Nr - the new 5g radio-access technology,'' in
  \emph{2018 IEEE 87th Vehicular Technology Conference (VTC Spring)}, 2018, pp.
  1--6.

\bibitem{IA3}
J.~Jeon, ``Nr wide bandwidth operations,'' \emph{IEEE Communications Magazine},
  vol.~56, no.~3, pp. 42--46, 2018.

\bibitem{IA4}
A.~A. Zaidi, R.~Baldemair, V.~Moles-Cases, N.~He, K.~Werner, and A.~Cedergren,
  ``Ofdm numerology design for 5g new radio to support iot, embb, and mbsfn,''
  \emph{IEEE Communications Standards Magazine}, vol.~2, no.~2, pp. 78--83,
  2018.

\bibitem{IA5}
X.~Lin, J.~Li, R.~Baldemair, J.-F.~T. Cheng, S.~Parkvall, D.~C. Larsson,
  H.~Koorapaty, M.~Frenne, S.~Falahati, A.~Grovlen, and K.~Werner, ``5g new
  radio: Unveiling the essentials of the next generation wireless access
  technology,'' \emph{IEEE Communications Standards Magazine}, vol.~3, no.~3,
  pp. 30--37, 2019.

\bibitem{SS3}
S.~Won and S.~W. Choi, ``A tutorial on 3gpp initial cell search: Exploring a
  potential for intelligence based cell search,'' \emph{IEEE Access}, vol.~9,
  pp. 100\,223--100\,263, 2021.

\bibitem{SS1}
A.~Chakrapani, ``On the design details of ss/pbch, signal generation and prach
  in 5g-nr,'' \emph{IEEE Access}, vol.~8, pp. 136\,617--136\,637, 2020.

\bibitem{SS2}
A.~Omri, M.~Shaqfeh, A.~Ali, and H.~Alnuweiri, ``Synchronization procedure in
  5g nr systems,'' \emph{IEEE Access}, vol.~7, pp. 41\,286--41\,295, 2019.

\bibitem{5GBook_IA}
\BIBentryALTinterwordspacing
E.~Dahlman, S.~Parkvall, and J.~Sköld, ``Chapter 16 - initial access,'' in
  \emph{5G NR: the Next Generation Wireless Access Technology}, E.~Dahlman,
  S.~Parkvall, and J.~Sköld, Eds.\hskip 1em plus 0.5em minus 0.4em\relax
  Academic Press, 2018, pp. 311--334. [Online]. Available:
  \url{https://www.sciencedirect.com/science/article/pii/B9780128143230000168}
\BIBentrySTDinterwordspacing

\bibitem{PRACH1}
Z.~Zhang, ``Novel prach scheme for 5g networks based on analog bloom filter,''
  in \emph{2018 IEEE Global Communications Conference (GLOBECOM)}, 2018, pp.
  1--7.

\bibitem{PRACH2}
G.~Schreiber and M.~Tavares, ``5g new radio physical random access preamble
  design,'' in \emph{2018 IEEE 5G World Forum (5GWF)}, 2018, pp. 215--220.

\bibitem{3GPP_211}
\BIBentryALTinterwordspacing
3GPP, ``{NR: Physical channels and modulation},'' {3rd Generation Partnership
  Project (3GPP)}, Technical Specification (TS) 36.211, 06 2022, version 17..0.
  [Online]. Available:
  \url{https://portal.3gpp.org/desktopmodules/Specifications/SpecificationDetails.aspx?specificationId=3213}
\BIBentrySTDinterwordspacing

\bibitem{3GPP_201}
\BIBentryALTinterwordspacing
------, ``{NR: Physical layer: General description},'' {3rd Generation
  Partnership Project (3GPP)}, Technical Specification (TS) 36.201, 06 2022,
  version 17.0.0. [Online]. Available:
  \url{https://portal.3gpp.org/desktopmodules/Specifications/SpecificationDetails.aspx?specificationId=3211}
\BIBentrySTDinterwordspacing

\bibitem{3GPP_213}
\BIBentryALTinterwordspacing
------, ``{NR: Physical layer procedures for control },'' {3rd Generation
  Partnership Project (3GPP)}, Technical Specification (TS) 36.213, 06 2022,
  version 17.2.0. [Online]. Available:
  \url{https://portal.3gpp.org/desktopmodules/Specifications/SpecificationDetails.aspx?specificationId=3215}
\BIBentrySTDinterwordspacing

\bibitem{LTE_SS}
Z.~Zhang, J.~Liu, and K.~Long, ``Low-complexity cell search with fast pss
  identification in lte,'' \emph{IEEE Transactions on Vehicular Technology},
  vol.~61, no.~4, pp. 1719--1729, 2012.

\bibitem{5GBook_NR}
\BIBentryALTinterwordspacing
E.~Dahlman, S.~Parkvall, and J.~Sköld, ``Chapter 5 - nr overview,'' in
  \emph{5G NR: the Next Generation Wireless Access Technology}, E.~Dahlman,
  S.~Parkvall, and J.~Sköld, Eds.\hskip 1em plus 0.5em minus 0.4em\relax
  Academic Press, 2018, pp. 57--71. [Online]. Available:
  \url{https://www.sciencedirect.com/science/article/pii/B9780128143230000053}
\BIBentrySTDinterwordspacing

\bibitem{5GBook_LTE}
\BIBentryALTinterwordspacing
------, ``Chapter 4 - lte—an overview,'' in \emph{5G NR: the Next Generation
  Wireless Access Technology}, E.~Dahlman, S.~Parkvall, and J.~Sköld,
  Eds.\hskip 1em plus 0.5em minus 0.4em\relax Academic Press, 2018, pp. 39--55.
  [Online]. Available:
  \url{https://www.sciencedirect.com/science/article/pii/B9780128143230000041}
\BIBentrySTDinterwordspacing

\bibitem{5GBook1}
M.~Enescu, Y.~Yuk, F.~Vook, K.~Ranta‐aho, J.~Kaikkonen, S.~Hakola, E.~Farag,
  S.~Grant, and A.~Manolakos, \emph{PHY Layer}, 2020, pp. 95--260.

\bibitem{5GBook_PHY}
\BIBentryALTinterwordspacing
S.~Ahmadi, ``Chapter 3 - new radio access physical layer aspects (part 1),'' in
  \emph{5G NR}, S.~Ahmadi, Ed.\hskip 1em plus 0.5em minus 0.4em\relax Academic
  Press, 2019, pp. 285--409. [Online]. Available:
  \url{https://www.sciencedirect.com/science/article/pii/B9780081022672000038}
\BIBentrySTDinterwordspacing

\bibitem{OFDM5}
J.~Bishop, J.-M. Chareau, and F.~Bonavitacola, ``Implementing 5g nr features in
  fpga,'' in \emph{2018 European Conference on Networks and Communications
  (EuCNC)}, 2018, pp. 373--9.

\bibitem{RFNOC1}
M.~Braun and J.~Pendlum, ``A flexible data processing framework for
  heterogeneous processing environments: Rf network-on-chip™,'' in \emph{2017
  International Conference on FPGA Reconfiguration for General-Purpose
  Computing (FPGA4GPC)}, 2017, pp. 1--6.

\bibitem{RFNOC2}
K.~M. Reddy, S.~J. Darak, and M.~D. Praveen, ``Novel framework for enabling
  hardware acceleration in gnu radio,'' in \emph{2020 IEEE International
  Symposium on Circuits and Systems (ISCAS)}, 2020, pp. 1--5.

\bibitem{5GPHY1}
F.~W. Vook, A.~Ghosh, E.~Diarte, and M.~Murphy, ``5g new radio: Overview and
  performance,'' in \emph{2018 52nd Asilomar Conference on Signals, Systems,
  and Computers}, 2018, pp. 1247--1251.

\bibitem{5GPHY2}
S.~Lagen, K.~Wanuga, H.~Elkotby, S.~Goyal, N.~Patriciello, and L.~Giupponi,
  ``New radio physical layer abstraction for system-level simulations of 5g
  networks,'' in \emph{ICC 2020 - 2020 IEEE International Conference on
  Communications (ICC)}, 2020, pp. 1--7.

\bibitem{5GLDPC1}
W.~Ji, Z.~Wu, K.~Zheng, L.~Zhao, and Y.~Liu, ``Design and implementation of a
  5g nr system based on ldpc in open source sdr,'' in \emph{2018 IEEE Globecom
  Workshops (GC Wkshps)}, 2018, pp. 1--6.

\bibitem{5GLDPC2}
P.~Henarejos and M.~Ángel Vázquez, ``Decoding 5g-nr communications via deep
  learning,'' in \emph{ICASSP 2020 - 2020 IEEE International Conference on
  Acoustics, Speech and Signal Processing (ICASSP)}, 2020, pp. 3782--3786.

\bibitem{5GLDPC3}
M.~Sybis, K.~Wesolowski, K.~Jayasinghe, V.~Venkatasubramanian, and
  V.~Vukadinovic, ``Channel coding for ultra-reliable low-latency communication
  in 5g systems,'' in \emph{2016 IEEE 84th Vehicular Technology Conference
  (VTC-Fall)}, 2016, pp. 1--5.

\bibitem{5GLDPC4}
H.~Gamage, N.~Rajatheva, and M.~Latva-aho, ``Channel coding for enhanced mobile
  broadband communication in 5g systems,'' in \emph{2017 European Conference on
  Networks and Communications (EuCNC)}, 2017, pp. 1--6.

\bibitem{5GOFDM1}
A.~Tewari, N.~Singh, S.~J.~D. amd V.~Kizheppatt, and M.~S. Jafri,
  ``Reconfigurable wireless phy with dynamically controlled out-of-band
  emission on zynq soc,'' in \emph{65TH IEEE INTERNATIONAL MIDWEST SYMPOSIUM ON
  CIRCUITS AND SYSTEMS (MWSCAS 2022)}, 2022, pp. 1--6.

\bibitem{5GOFDM2}
L.~Zhang, A.~Ijaz, P.~Xiao, M.~M. Molu, and R.~Tafazolli, ``Filtered ofdm
  systems, algorithms, and performance analysis for 5g and beyond,'' \emph{IEEE
  Transactions on Communications}, vol.~66, no.~3, pp. 1205--1218, 2018.

\bibitem{5GBF1}
W.~Hong, J.~Choi, D.~Park, M.-s. Kim, C.~You, D.~Jung, and J.~Park, ``mmwave 5g
  nr cellular handset prototype featuring optically invisible beamforming
  antenna-on-display,'' \emph{IEEE Communications Magazine}, vol.~58, no.~8,
  pp. 54--60, 2020.

\bibitem{5GBF2}
\BIBentryALTinterwordspacing
C.~Herranz, M.~Zhang, M.~Mezzavilla, D.~Martin-Sacrist\'{a}n, S.~Rangan, and
  J.~F. Monserrat, ``A 3gpp nr compliant beam management framework to simulate
  end-to-end mmwave networks,'' in \emph{Proceedings of the 21st ACM
  International Conference on Modeling, Analysis and Simulation of Wireless and
  Mobile Systems}, ser. MSWIM '18.\hskip 1em plus 0.5em minus 0.4em\relax New
  York, NY, USA: Association for Computing Machinery, 2018, p. 119–125.
  [Online]. Available: \url{https://doi.org/10.1145/3242102.3242117}
\BIBentrySTDinterwordspacing

\bibitem{5GCE1}
M.~Mehrabi, M.~Mohammadkarimi, M.~Ardakani, and Y.~Jing, ``Decision directed
  channel estimation based on deep neural network $k$ -step predictor for mimo
  communications in 5g,'' \emph{IEEE Journal on Selected Areas in
  Communications}, vol.~37, no.~11, pp. 2443--2456, 2019.

\bibitem{5GCE2}
H.~He, C.-K. Wen, S.~Jin, and G.~Y. Li, ``Deep learning-based channel
  estimation for beamspace mmwave massive mimo systems,'' \emph{IEEE Wireless
  Communications Letters}, vol.~7, no.~5, pp. 852--855, 2018.

\bibitem{5GCE3}
C.-J. Chun, J.-M. Kang, and I.-M. Kim, ``Deep learning-based channel estimation
  for massive mimo systems,'' \emph{IEEE Wireless Communications Letters},
  vol.~8, no.~4, pp. 1228--1231, 2019.

\bibitem{5Gsplit1}
L.~M.~P. Larsen, A.~Checko, and H.~L. Christiansen, ``A survey of the
  functional splits proposed for 5g mobile crosshaul networks,'' \emph{IEEE
  Communications Surveys \& Tutorials}, vol.~21, no.~1, pp. 146--172, 2019.

\bibitem{5Gsplit2}
I.~Koutsopoulos, ``The impact of baseband functional splits on resource
  allocation in 5g radio access networks,'' in \emph{IEEE INFOCOM 2021 - IEEE
  Conference on Computer Communications}, 2021, pp. 1--10.

\bibitem{5Gsplit3}
R.~I. Rony, E.~Lopez-Aguilera, and E.~Garcia-Villegas, ``Optimization of 5g
  fronthaul based on functional splitting at phy layer,'' in \emph{2018 IEEE
  Global Communications Conference (GLOBECOM)}, 2018, pp. 1--7.

\bibitem{ofdm2}
B.~{Drozdenko}, M.~{Zimmermann}, {Tuan Dao}, M.~{Leeser}, and K.~{Chowdhury},
  ``High-level hardware-software co-design of an 802.11a transceiver system
  using zynq soc,'' in \emph{2016 IEEE Conference on Computer Communications
  Workshops (INFOCOM WKSHPS)}, 2016, pp. 682--683.

\bibitem{ofdm3}
B.~{Drozdenko}, M.~{Zimmermann}, T.~{Dao}, K.~{Chowdhury}, and M.~{Leeser},
  ``Hardware-software codesign of wireless transceivers on zynq heterogeneous
  systems,'' \emph{IEEE Transactions on Emerging Topics in Computing}, vol.~6,
  no.~4, pp. 566--578, 2018.

\bibitem{OFDM4}
S.~{Garg}, N.~{Agrawal}, S.~J. {Darak}, and P.~{Sikka}, ``Spectral coexistence
  of candidate waveforms and dme in air-to-ground communications: Analysis via
  hardware software co-design on zynq soc,'' in \emph{2017 IEEE/AIAA 36th
  Digital Avionics Systems Conference (DASC)}, 2017, pp. 1--6.

\bibitem{ofdm1}
T.~H. {Pham}, S.~A. {Fahmy}, and I.~V. {McLoughlin}, ``An end-to-end
  multi-standard ofdm transceiver architecture using fpga partial
  reconfiguration,'' \emph{IEEE Access}, vol.~5, pp. 21\,002--21\,015, 2017.

\end{thebibliography}

\end{document}